\documentclass[prb,preprint]{revtex4-1} 

\usepackage{amsmath}  
\usepackage{amsfonts} 
\usepackage{graphicx} 
\usepackage{floatrow}
\usepackage{siunitx}
\newfloatcommand{capbtabbox}{table}[][\FBwidth]
\usepackage{blindtext}
\usepackage{setspace}
\usepackage{listings}
\usepackage{upquote}
\usepackage{xcolor}
\usepackage{caption}
\usepackage{placeins}

\usepackage{subcaption}
\usepackage{lipsum}
\DeclareFloatSeparators{myfill}{\hskip.013\textwidth plus1fill}
\DeclareFloatVCode{largevskip}%
{\vskip 10pt}
\DeclareCaptionFont{white}{\color{white}}
\DeclareCaptionFormat{listing}{%
  \parbox{\textwidth}{\colorbox{gray}{\parbox{\textwidth}{#1#2#3}}\vskip-4pt}}
\def\ContinueLineNumber{\lstset{firstnumber=last}}
\def\StartLineAt#1{\lstset{firstnumber=#1}}

\begin{document}
\belowcaptionskip=-5pt


\title{Combining high-performance hardware, cloud computing, and deep learning frameworks to accelerate physical simulations: probing the Hopfield network}

\author{Vaibhav S. Vavilala}
\email{vsv2109@columbia.edu} 
\affiliation{Department of Computer Science, Columbia University, New York, NY 10027}



\date{\today}

\begin{abstract}

The synthesis of high-performance computing (particularly graphics processing units), cloud computing services (like Google Colab), and high-level deep learning frameworks (such as PyTorch) has powered the burgeoning field of artificial intelligence. While these technologies are popular in the computer science discipline, the physics community is less aware of how such innovations, freely available online, can improve research and education. In this tutorial, we take the Hopfield network as an example to show how the confluence of these fields can dramatically accelerate physics-based computer simulations and remove technical barriers in implementing such programs, thereby making physics experimentation and education faster and more accessible. To do so, we introduce the cloud, the GPU, and AI frameworks that can be easily repurposed for physics simulation. We then introduce the Hopfield network and explain how to produce large-scale simulations and visualizations for free in the cloud with very little code (fully self-contained in the text). Finally, we suggest programming exercises throughout the paper, geared towards advanced undergraduate students studying physics, biophysics, or computer science.

\end{abstract}

\maketitle 

\section{Introduction} 

Computation is critical to calculate physical properties of modeled systems. Some experiments are impossible to perform but can be simulated if the theory known and sufficient computational tools are available. Concepts that feel abstract to students can become tangible when simulated on a computer. Although computation is a component of most undergraduate physics curricula, it is commonly under-emphasized, and educators must keep pace with the rapidly improving technology to produce the most capable students.\cite{doi:10.1119/1.4837437} In this paper, we discuss innovations that have improved accessibility to numerical computations (via the cloud), and accelerated the performance of large-scale simulations (via the graphics processing unit). 

Frequently the bottleneck of numerical computations involves matrix operations. Such examples include matrix diagonalization to solve the Schr\"{o}dinger Equation and solving systems of linear or differential equations to perform Finite Element Analysis. These computations are generally of $O(N^3)$ complexity (i.e. scaling the dimensions of the input matrix by $10\times$ results in a $1000\times$ performance hit). In the absence of high-performance hardware, most numerical computations cannot scale well, which slows if not entirely precluding simulations beyond a certain size. To perform simulations, students commonly use scripting languages such as \textsc{Matlab} or Python in the classroom and laboratory. Using a personal machine involves an up-front cost of programming environment setup, and the speed of simulations is limited by the hardware (typically two or four CPUs). Workstations with high-performance hardware are available only in the most well-funded labs. To help address limitations in the computational aspects of current physics curricula, we show that the software and hardware innovations that have accompanied the rise of deep learning can dramatically enhance the runtime performance and accessibility of computational physics. We start with a brief introduction to deep learning.

The past decade has seen a meteoric rise in artificial intelligence research, and the bulk of this progress is from deep learning - which aims to solve complex problems by constructing computational models that learn from a high volume of data.\cite{Goodfellow:2016:DL:3086952} For example, deep learning models can classify thousands of objects in images with very little error, synthesize speech from text, and detect anomalies in financial transactions to stop fraud. Although the key algorithm for training deep learning models - backpropagation - has been known since 1986,\cite{Rumelhart:1986we} it was only in the past decade or so that researchers recognized the value of collecting vast amounts of data, and corporations such as NVIDIA created hardware that could feasibly train large-scale deep learning models via the graphics processing unit (GPU).\cite{pmlr-v28-coates13} From personal computing to industrial data centers, most computers hold between 2 - 96 CPU cores. A single GPU has embedded within it thousands of processing cores - which, although each GPU core individually is slower than a CPU core, the sheer quantity of cores in a GPU enables it to dwarf the performance of multiple CPU cores in tasks that can be parallelized. For example, when performing matrix multiplication, each element of the product can be computed independently and in parallel. Hence for large matrices, matrix multiplication can be performed significantly faster on a GPU than on a CPU. In fact, several problems can benefit from the GPU such as backpropagation (in which gradients of millions of variables can be computed in parallel), rendering\cite{Akenine-Moller:2008:RR:2829183} (in which thousands of pixel colors on a screen can be computed in parallel), and the simulation of natural phenomena like fluids,\cite{DBLP:journals/corr/abs-1806-02071} cloth,\cite{kavan2011physics} \& hair\cite{Tariq2008RealTH} (which are commonly GPU-accelerated in the film and games industries). A typical simulation will involve the GPU(s) and CPU(s) working together, whereby general-purpose instructions (like loading data or plotting graphs) run on the CPU, and parallelizable, compute-intensive parts of the application (like matrix calculations) run on the GPU.

While the GPU enjoys several use-cases, initially only researchers with a computer science background could access its benefits, as strong knowledge of a low-level programming language like CUDA would be required. With the open-sourcing of deep learning frameworks such as PyTorch\cite{NIPS2019_9015} and TensorFlow,\cite{tensorflow2015-whitepaper} physicists no longer face such programming language barriers. These frameworks allow developers to code in a high-level scripting language (usually Python) and abstract away low-level CUDA function calls. The mathematical underpinnings of most physics simulations including sampling, linear algebra, numerical optimization, signal processing, and differentiation are implemented in these frameworks in an aggressively optimized manner, using \textsc{Matlab}/NumPy-like syntax that is intuitive to non-programmers. Additionally, these frameworks are supported by extensive function manuals and numerous high-quality courses such as fast.ai. The physicist, as a result, reaps the benefits of the rapidly improving technology and can remain focused on the scientific aspects of the simulation, instead of implementation details. 

Furthermore, the advantage of using frameworks that garner strong adoption like PyTorch and TensorFlow is that with greater usage, bugs are caught sooner, and robust documentation \& support are further justified. Crucially, with more users, online programming help communities like StackOverflow become filled with questions \& solutions to common (and uncommon) programming errors, aiding physics students and researchers in rapidly fixing their own bugs by benefitting from crowdsourced knowledge.

Yet obstacles remain - GPUs cost hundreds to thousands of dollars, often impractical for students and schools in developing countries. Further, installing an optimized deep learning framework with GPU support is non-trivial and hardware-specific, often requiring several days to complete even for experts. Although most personal computers on sale today ship with a GPU, they are typically limited in memory, and the user still runs into environment setup challenges. Cloud computing solves these problems. With high-end GPUs and pre-built deep learning environments available over the Internet, anyone can write simulations on their personal computer, remotely execute their code on a cloud-based machine, and visualize the results in real-time.  

In the remainder of this tutorial, we first show how to set up a GPU instance in the cloud, pre-loaded with PyTorch. We then introduce the foundations of the Hopfield neural network (HNN) - including its theoretical roots in condensed matter physics and its applications in AI. From there, we guide the reader through simulating the HNN on the GPU with few lines of code, showing the drastic improvement in performance as compared with the CPU. We conclude with suggested programming exercises to reproduce famous results of the HNN.

\section{Colab \& PyTorch Environment Setup}

For this article, we take Google Colab\cite{8485684} as our (currently free) cloud provider of choice, although Kaggle, Azure Notebooks, Paperspace Gradient, and Amazon Sagemaker are among the alternatives we are aware of. We use the PyTorch deep learning library here, and note that TensorFlow is an outstanding alternative. CuPy~\cite{okuta2017cupy} is a package for purely numerical computing that is also worth considering. Our hardware of choice are NVIDIA GPUs, as these chips benefit from the strongest support in the deep learning community. We note that certain matrix operations can be further accelerated with Tensor Processing Units~\cite{8358031} (TPUs), freely available on Colab, but such hardware is outside the scope of this paper.

To access Colab, navigate to \href{https://colab.research.google.com}{https://colab.research.google.com}. Sign in with a Google account, and then click \textit{connect} to request an instance. To obtain a GPU, click \textit{Runtime} $\rightarrow$ \textit{Change Instance Type} $\rightarrow$  \textit{GPU} for Hardware Accelerator. In our experience, GPUs are available immediately upon request. The user then has access to a pre-built Python environment with PyTorch, TensorFlow, and common scientific computing packages like NumPy, SciPy, and Matplotlib pre-installed. Additional packages from GitHub or the Python Package Index can easily be installed in-browser to augment the pre-built environment. For example, a biophysics student studying protein interactions may wish to use the pypdb package. Installing this package is as simple as \texttt{!pip install pypdb}. Within Colab, the user can write Python code in-browser using a robust Integrated Development Environment (IDE) with tab completion and code formatting. All code is automatically backed-up in Google Drive, which allows for easy code sharing/collaboration and the ability to work from any machine connected to the Internet. We note that although the details of how to access a particular cloud provider will change over time and based on which service provider is used, the concept of leveraging this trifecta of technologies (HPC, Cloud, AI) will become increasingly important in performing computer simulations in the years to come. We further note that the present article focuses on using GPU-accelerated numerical libraries contained in deep learning frameworks, as opposed to using deep learning itself in physics research. Although AI algorithms have started to enjoy a symbiotic relationship with data-heavy physical simulations,\cite{baldi2014searching,mehta2019high,schütt2018schnetpack} such applications are beyond the scope of this paper. 


We can verify the environment setup with the following example. In Code Sample 1, we first import the PyTorch and timing packages. We then construct a large random matrix $A_{16000\times16000}$ sampled from a uniform distribution $A_{ij}\in[0,1]$ on the specified hardware. Note that the \texttt{device} variable in line 3 defines whether a tensor should be stored on the GPU (\texttt{device="cuda"}) or CPU (\texttt{device="cpu"}). We finally compute the matrix inverse in line 5:

\singlespacing
\begin{lstlisting}[label=env_setup,caption=Test PyTorch environment (matrix inverse)]
import torch, time
start_time = time.time() #Begin timer
device='cuda' #'cpu' for CPU
A=torch.rand(16000,16000, device=device)
iA = A.inverse()
print('execution time: ' + str(time.time() - start_time))
\end{lstlisting}
\setstretch{2}

The code executes in 2.98 seconds on the GPU and 53.4 seconds on the CPU. To show the ease of performing matrix multiplication, we take an example use-case where a modeled linear system has more unknowns than observations, and a least-squares solution is desired. We construct a large random matrix $A_{8000\times10000}$ sampled from a uniform distribution $A_{ij}\in[0,1]$ on the specified hardware (GPU or CPU), and initialize a random vector $b_{8000\times1}$. To solve the underdetermined linear system $Ax=b$, we use the Moore-Penrose pseudoinverse in line 13, $x=(A^{\top}A)^{-1}A^{\top}b$. 

\StartLineAt{7}
\singlespacing
\begin{lstlisting}[label=env_setup,caption=Test PyTorch environment (pseudoinverse)]
import torch, time
start_time = time.time() #Begin timer
device='cuda' #'cpu' for CPU
A=torch.rand(8000,10000, device=device)
b=torch.rand(8000,1,device=device)
A_t = torch.t(A) #precompute matrix transpose
x=torch.matmul(torch.matmul(A_t,A).inverse(),torch.matmul(A_t,b))
print('execution time: ' + str(time.time() - start_time))
\end{lstlisting}
\setstretch{2}

On the GPU, the code executes in 1.46 seconds. On the CPU, it requires 45.1 seconds. We encourage the reader to reproduce our timings, and observe that as matrix inversion and matrix multiplication are of $O(N^3)$ complexity, the timings scale as such. These brief examples show the order of magnitude superior performance of the GPU in scientific computing, and its ease of access. As an exercise, generate a random square matrix sampled from a Gaussian distribution and diagonalize the matrix. Plot how the performance scales with the input size, and compare the CPU vs. GPU timings. You can use the PyTorch documentation to obtain the syntax and the Matplotlib package for visualization.

To illustrate the value of the GPU in studying physical systems, we take the Hopfield network as a specific example. In the next section, we introduce the theory behind the HNN.

 \section{The Hopfield network}
The Hopfield neural network is a two-state information processing model first described in 1982.\cite{hop} The dynamical system exhibits numerous physical properties relevant to the study of spin glasses (disordered magnets),\cite{dja1} biological neural networks,\cite{baryam} and computer science (including circuit design,~\cite{crct} the traveling salesman problem,~\cite{tsp} image segmentation,~\cite{imseg} and character recognition\cite{Widodo_2018}). 

\begin{figure}[h!]
\centering
\includegraphics{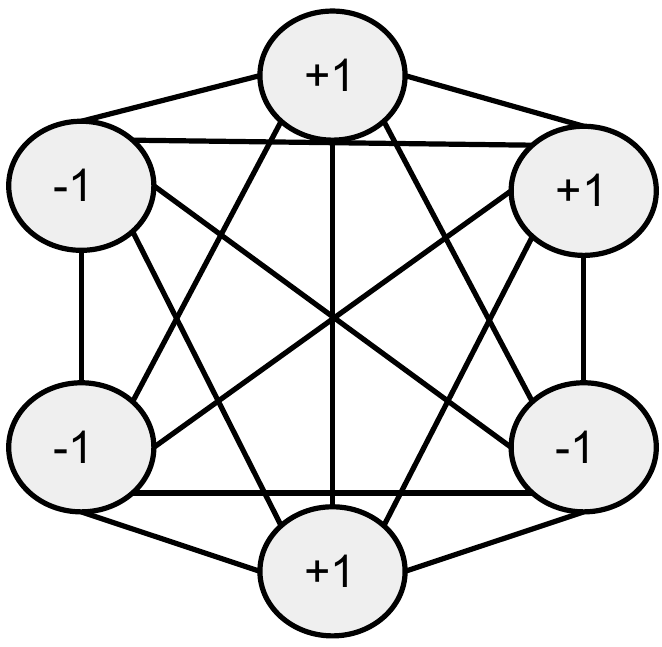}
\caption{A fully-connected Hopfield neural network with $N=6$ and instantaneous state $S=\{+1,+1,-1,+1,-1,-1\}$, starting from the top and indexing clockwise. The synaptic weights are stored in the $6\times6$ matrix $J_{ij}$.}
\label{NNDiag}
\end{figure}

The network consists of $N$ fully-connected neurons that can store $N$-tuples of $\pm1$'s (shown in Fig.~\ref{NNDiag}). Each pair of neurons has an associated weight stored in a symmetric synaptic weight matrix $J_{ij}$ with $i,j \in 1...N$. The binary state of a neuron represented by $S_i$ is mapped onto a classical Ising spin,\cite{mbf} where $S_i=+1$ $(-1)$ represents a neuron that is firing (at rest). In the binary representation, such a neuron fires when its potential exceeds a threshold $U_i$ that is independent of the state $S_i$ of the neuron. The state of the neuron at time $t+1$ is determined solely by the sum total of post-synaptic potential contributions from all other neurons at time $t$. This assumption, where the time evolution of a neural state is determined by the local field produced by other neurons, allows us to associate a classical Hamiltonian~\cite{mbf} (or energy functional) consistent with the discrete, asynchronous time evolution of the neural network. The neural network is, thus, mapped onto an Ising model with long-ranged, generically frustrated interaction. Spin glass approaches have been highly fruitful in investigating the properties of such Hopfield neural networks near criticality.~\cite{dja1} In particular, they have quantified the critical memory loading $\alpha_c\sim 0.144$, such that when $p$ patterns are imprinted on a network with $N$ neurons, the network has no faithful retrieval for $\alpha=p/N>\alpha_c$ whereas when $\alpha<\alpha_c$, the memory retrieval is accompanied by a small fraction of error (for example, no more than 1.5\% of bits flipped).~\cite{dja1}

We consider a network of $N$ two-state neurons $S_i=\pm 1$ trained with $p=\alpha N$ random patterns $\xi^\mu_k$ where $\mu=1,\ldots, p$, $k=1,\ldots, N$, and $N \gg 1$. The symmetric synaptic matrix is created from the $p$ quenched patterns,
\begin{equation}
\label{eq:jij}
J_{ij}=\frac{1}{N}\sum_{\mu=1}^p \xi^\mu_i\xi^\mu_j= J_{ji}\hspace{1cm} i\neq j.
\end{equation}

With symmetric neural interconnections, a stable state is ultimately reached.\cite{1057328} Further, we note that the neural interconnections are considered fixed after training, and $J_{ii}=0$ in the traditional interpretation. However, previous work has shown that the hysteretic (or self-interaction) terms enhance retrieval quality, especially in the presence of stochastic noise.\cite{GOPALSAMY2007375,PhysRevE.64.051912,tsuboshita2010statistical,bharitkar2000hysteretic} Hysteresis is a property found in biological neurons (via a refractory period after a neuron fires) and is inherent in many physical, engineering, and economic systems. Thus, in this paper we set $J_{ii}=\lambda\alpha$ to probe the effects of self-action, with $\lambda=0$ representing the traditional model.

We compute the simulated memory capacity by flipping some fraction of bits less than a Hamming distance $N/2$ away from each imprinted pattern, time evolving the corrupted probe vectors through the network until some convergence criterion, and then measuring the quality of recall. The Hamming distance is defined as the number of different bits between two patterns. The zero-temperature network dynamics are as follows:

\begin{equation}
S_i(t+1)=\mathrm{sign}\left[\sum_{j\neq i}^{N} J_{ij} S_j(t)\right]. 
\label{eq:time}
\end{equation}

The network evolving under this deterministic update rule behaves as a thermodynamical system in such a way as to minimize an overall energy measure defined over the whole network.\cite{McClelland1988} These low energy states are called attractor states. When $\alpha<\alpha_c$, the imprinted patterns are the attractors. Above criticality, non-imprinted local minima, called spurious memories, also become dynamically stable states. The basin of attraction is defined as the maximal fraction of bits that can be flipped such that the probe vector still relaxes to its intended imprint within a small fraction of error.\cite{baryam} At low memory loading, the basin of attraction of imprinted patterns is very high, near $N/2$, and beyond criticality $\alpha_c$, the basin of attraction vanishes. We let $m_0$ be the normalized dot-product between an imprinted pattern and its corrupted probe vector, and let $m_f$ represent the final overlap between a time-evolved probe vector and its intended imprint. Therefore, an imprint with $0.1N$ bits flipped would have an overlap of $m_0 = 0.8$. We note that asynchronous update refers to each neuron updating serially (in random order, per time step), and synchronous update refers to all neurons updated at once per time step.

Despite the simplicity of the Hopfield network, considerable computational power is inherent in the system. We find numerous interesting properties like Hebbian learning, associative recall (whereby similar to the human brain, whole memories can be recovered from parts of them), and robustness to thermal and synaptic noise.\cite{baryam} 

In the next section, we demonstrate how to simulate the Hopfield network on the GPU with a large system size ($N=32K$) and very few lines of code.

\section{Implementing Simulations on the GPU}

Here we show with code how to implement a Hopfield network simulation. After setting up a PyTorch environment in Google Colab as specified in the Introduction, we start by defining functions to construct the set of imprinted memories, the synaptic weight matrix (as defined in eq. \ref{eq:jij}), and probe vectors perturbed from the imprints.  

\StartLineAt{15}
\singlespacing
\begin{lstlisting}[label=env_setup,caption=Initialize synaptic weight matrix]
def initSynapticMatrix(N, V, L):
    """Constructs the synaptic weight matrix

    Args:
      N:      number of neurons
      V:      N x p matrix containing set of imprinted memories
      L:      the diagonal term
    Result:
      J:      the N x N synaptic matrix
    """

    J=(1.0/N)*torch.matmul(V,torch.t(V)) 
    # set the diagonal self-action terms with L
    J.as_strided([N],[N + 1]).copy_(torch.diag(J)*L)
    return J
\end{lstlisting}
\setstretch{2}

\StartLineAt{30}
\singlespacing
\begin{lstlisting}[label=env_setup,caption=Perturb probe vectors]
def flipBits(m0, N, p, device, probe):
    """Randomly negates elements of the input.

    Args:
      m0:     desired dot product overlap after flipping bits
      N:      number of neurons
      p:      number of patterns
      device: cpu or gpu
      probe:  N x p probe matrix whose elements should be flipped
    """

    #corrupt (1-m0)*N/2 random bits per pattern
    nFlip=round((1-m0)*N/2) #How many bits to flip for overlap m0
    if nFlip > 0:
        #random sample indices from every vector
        y=torch.multinomial(torch.ones(p,N,device=device),nFlip) 
        #corresponding index to bit flip
        r=torch.arange(0,p,1,device=device).expand(nFlip,p) 
        #flip bits in probe array
        probe[y.reshape(-1),torch.t(r).reshape(-1)]*=-1 
\end{lstlisting}
\setstretch{2}

\StartLineAt{50}
\singlespacing
\begin{lstlisting}[label=env_setup,caption=Time evolve network]
def evolveNetwork(N, probe, probe_new, J, V, t_type, dotp_evol, num_conv, device):
    """Time evolves the Hopfield network.

    Args:
      N:         number of neurons
      probe:     N x p probe matrix of imprints
      probe_new: N x p probe matrix of imprints after one time step
      J:         N x N synapic weight matrix
      V:         N x p matrix containing set of imprinted memories
      t_type:    torch.FloatTensor if CPU, torch.cuda.FloatTensor if GPU
      dotp_evol: max_steps+1 x p matrix storing each pattern's dot product overlap per time step
      num_conv:  1 x max_steps+1 matrix storing the number of converged states per time step
      device:    CPU or GPU
    """
    #keep track of non-converged pattern indices
    nidx = torch.arange(0,p,1,device=device)
    for i in range(1,max_steps+1):
        probe_new[:,nidx] = torch.matmul(J,probe[:,nidx])
        #add noise to zero elements and take sign function
        probe_new[:,nidx] = torch.sign(probe_new[:,nidx]+(probe_new[:,nidx]==0).type(t_type)*(2*torch.rand(N,len(nidx),device=device)-1))
        dotp_evol[i,:]=torch.sum(probe_new*V,dim=0)
        nidx = nidx[torch.sum(probe_new[:,nidx]*probe[:,nidx],dim=0)!=N]
        num_conv[0,i]=nidx.nelement()
        if nidx.nelement() == 0:
          dotp_evol[i+1:,:] = dotp_evol[i,:]
          num_conv[0,i+1:] = num_conv[0,i]
          print('converging early: ' + str(i) + ' tsteps to converge')
          break
        probe=probe_new.clone()
\end{lstlisting}
\setstretch{2}

We then import key packages in line 79 and specify that our code should run on the GPU. The same code can be run on the GPU or CPU by simply switching the \texttt{dst} variable between \texttt{"cuda"} and \texttt{"cpu"} in line 80.  In subsequent lines, we define the system parameters including the network size $N$, memory loading $\alpha$, initial overlap $m_0$, and self-interaction term $\lambda$. 

\StartLineAt{79}
\singlespacing
\begin{lstlisting}[label=params,caption=Define system parameters]
import torch, numpy as np, time
dst = 'cuda'  #cpu for cpu-only mode
t_type=torch.FloatTensor
if dst=='cuda':
    torch.cuda.empty_cache() #frees memory for large matrices
    t_type = torch.cuda.FloatTensor
device = torch.device(dst)
start_time = time.time() #Begin timer
#System parameters
N_list=[1000,4000,16000]
alpha=0.24 #memory loading
L=1 #diagonal coefficient lambda
m0=0.9 #initial overlap
max_steps = 100 #max tsteps
dotp_lists = [] #track the evolution of the overlap m over time
num_conv_lists = [] #track the number of converged states over time
\end{lstlisting}
\setstretch{2}

\ContinueLineNumber

Using these parameters, we are ready to construct the set of imprinted memories (line 100), the synaptic weight matrix (line 103, as defined in eq. \ref{eq:jij}), and probe vectors perturbed from the imprints (line 107). Finally, the probe vectors are repeatedly updated according to eq. \ref{eq:time} until either convergence or the max time steps are reached (line 121).

\StartLineAt{95}
\singlespacing
\begin{lstlisting}[label=init,caption=Execute simulation]
for N in N_list:
    print('Running simulation for N=' + str(N))
    p=int(alpha*N) #number of imprints

    #Construct set of patterns
    V=2*torch.round(torch.rand(N,p, device=device))-1  

    #Construct synaptic matrix including diagonal terms
    J = initSynapticMatrix(N,V,L)
    
    #corrupt patterns such that initial overlap is m0
    probe = V.clone() #Construct a probe matrix
    flipBits(m0, N, p, device, probe)

    #Initialize variables for storing simulation data
    probe_new = probe.clone() 
    
    #store time evolution of dot-products
    dotp_evol = torch.zeros(max_steps+1,p,device=device) 
    dotp_evol[0,:] = torch.sum(probe*V,dim=0) # time t=0

    #keep track of how many states have not converged over time
    num_conv = torch.zeros(1,max_steps+1,device=device) 
    num_conv[0,0]=p #At time t=0, p states have not converged
    
    #main neural updating loop: time evolve network until convergence
    evolveNetwork(N, probe, probe_new, J, V, t_type, dotp_evol, num_conv, device)

    #store the dot product evolutions and number of converged states  
    dotp_lists.append((dotp_evol/N).cpu())
    num_conv_lists.append((p-num_conv.cpu())/p) 

print('execution time: ' + str(time.time() - start_time))
  \end{lstlisting}
\setstretch{2}
\ContinueLineNumber

\begin{figure}[t]
\setlength\fboxsep{0pt}\setlength\fboxrule{0.75pt}
\ffigbox[\textwidth]
{
\begin{subfloatrow}[2]
\ffigbox[.49\textwidth]
  {
    \caption{table comparing timings}
    \label{subfig:tabTime}
  }
  {
      \begin{tabular}{ccc} \hline
  N & GPU (sec)  & CPU (sec) \\ \hline
  $1K$ & \SI{6.75e-2} & $3.76\times10^{-1}$ \\
  $2K$ & \SI{1.33e-1} & $2.54$ \\
  $4K$ & \SI{4.19e-1} &  $2.07\times10^1$ \\
  $8K$ & \SI{2.41e0} &  $1.64\times10^2$ \\
  $16K$ & \SI{1.86e1} &  $1.44\times10^3$ \\
  $32K$ & \SI{2.09e2} &  $1.15\times10^4$* \\\hline \\\\\\
  \end{tabular}
  }
\ffigbox[.49\textwidth]
  {
    \caption{plot of the timings}
    \label{subfig:plotGPU}
  }
  {
    \includegraphics[width=\linewidth]{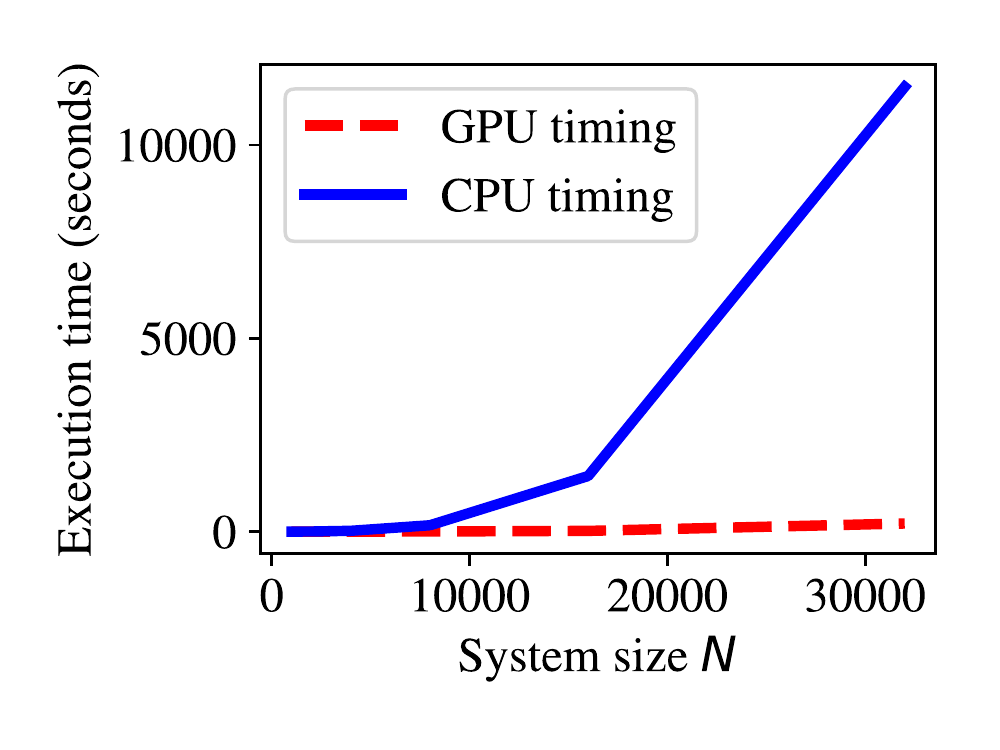}%
  }
\end{subfloatrow}%
}
{
    \caption{Comparing the performance of Hopfield network simulation on the GPU vs. CPU. For large $N$, the GPU version executes 50-80 times faster than the CPU.  Note that the CPU timing for $N=32K$ is an estimate based on cubic scaling and not actually simulated.}
    \label{fig:gpucpu}
}
\end{figure}%

With our code complete, we are ready to demonstrate the advantages of using the GPU over CPU. In fig.~\ref{fig:gpucpu} we compare timings for a simulation with $(\alpha,\lambda,m_0,max\_steps)=(0.12,0,0.8,100)$. For CPU timings, we use two Intel Xeon CPUs @ 2.20GHz. For GPU timings, we use one NVIDIA Tesla T4 GPU with 15GB allocated RAM. In the author's experience, such a GPU has consistently been available in the USA. For large $N$, the GPU-accelerated simulations execute 50-80 times faster than the CPU mode. As the bottleneck of Hopfield network simulation is matrix multiplication, we observe that the asymptotic complexity scales $O(N^3)$ with the input size.

\section{Visualizations \& Extensions}

\begin{figure}[t]

\setlength\fboxsep{0pt}\setlength\fboxrule{0.75pt}
\ffigbox[\textwidth]
{
\begin{subfloatrow}[2]
\ffigbox[.45\textwidth]
  {
    \caption{$P(m)$ histogram with $\alpha=0.13<\alpha_c$}
    \label{subfig:Lam0Pm0}
  }
  {
\includegraphics[width=\linewidth]{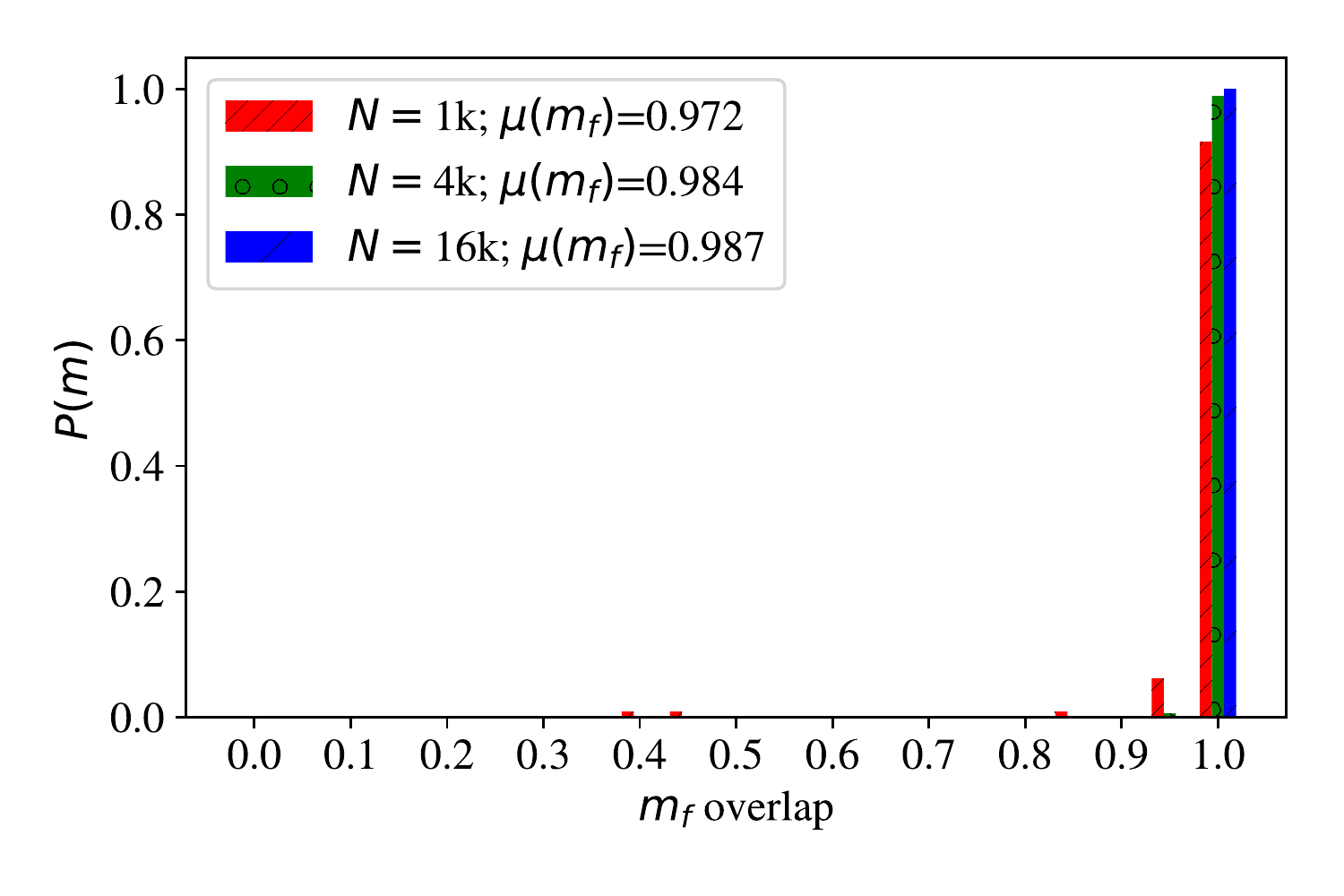}%
  }
\ffigbox[.45\textwidth]
  {
    \caption{$P(m)$ histogram with $\alpha=0.15>\alpha_c$}
    \label{subfig:Lam0Pm1}
  }
  {
    \includegraphics[width=\linewidth]{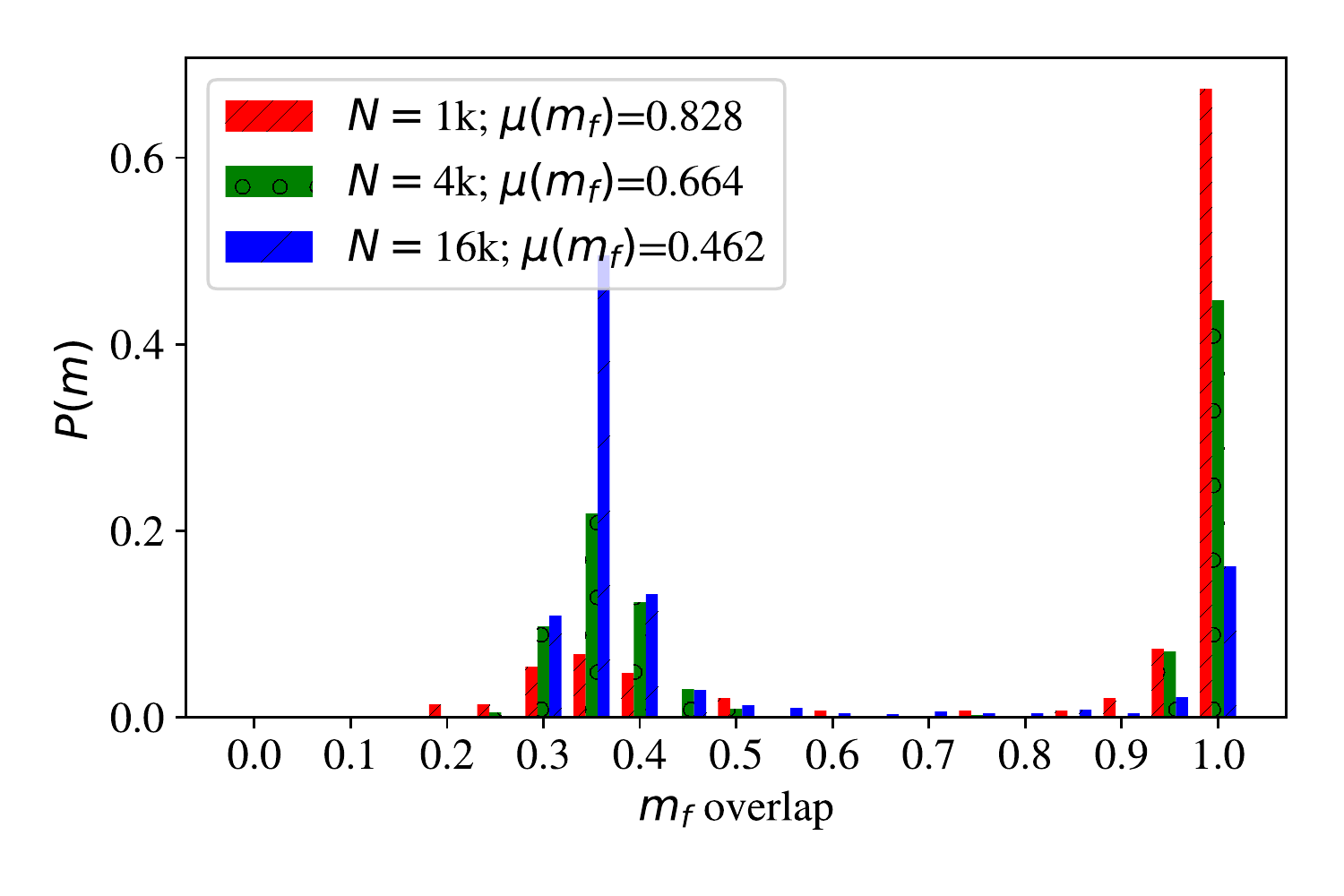}%
  }

\end{subfloatrow}

\begin{subfloatrow}[2]
\ffigbox[.45\textwidth]
  {
    \caption{pattern convergence with $\alpha=0.13<\alpha_c$}
    \label{subfig:Lam0Frac0}
  }
  {
\includegraphics[width=\linewidth]{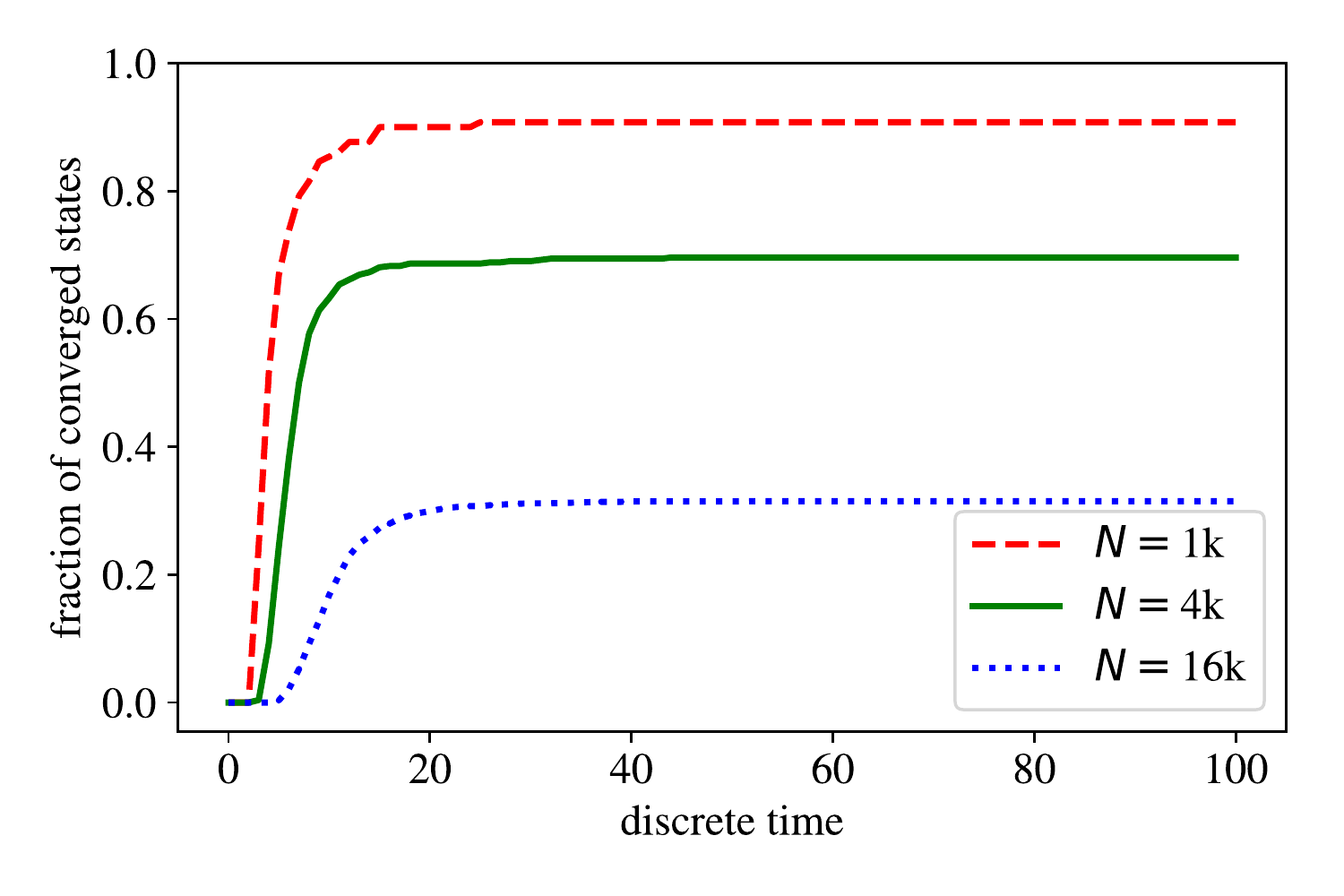}%
  }
\ffigbox[.45\textwidth]
  {
    \caption{pattern convergence with $\alpha=0.15>\alpha_c$}
    \label{subfig:Lam0Frac1}
  }
  {
    \includegraphics[width=\linewidth]{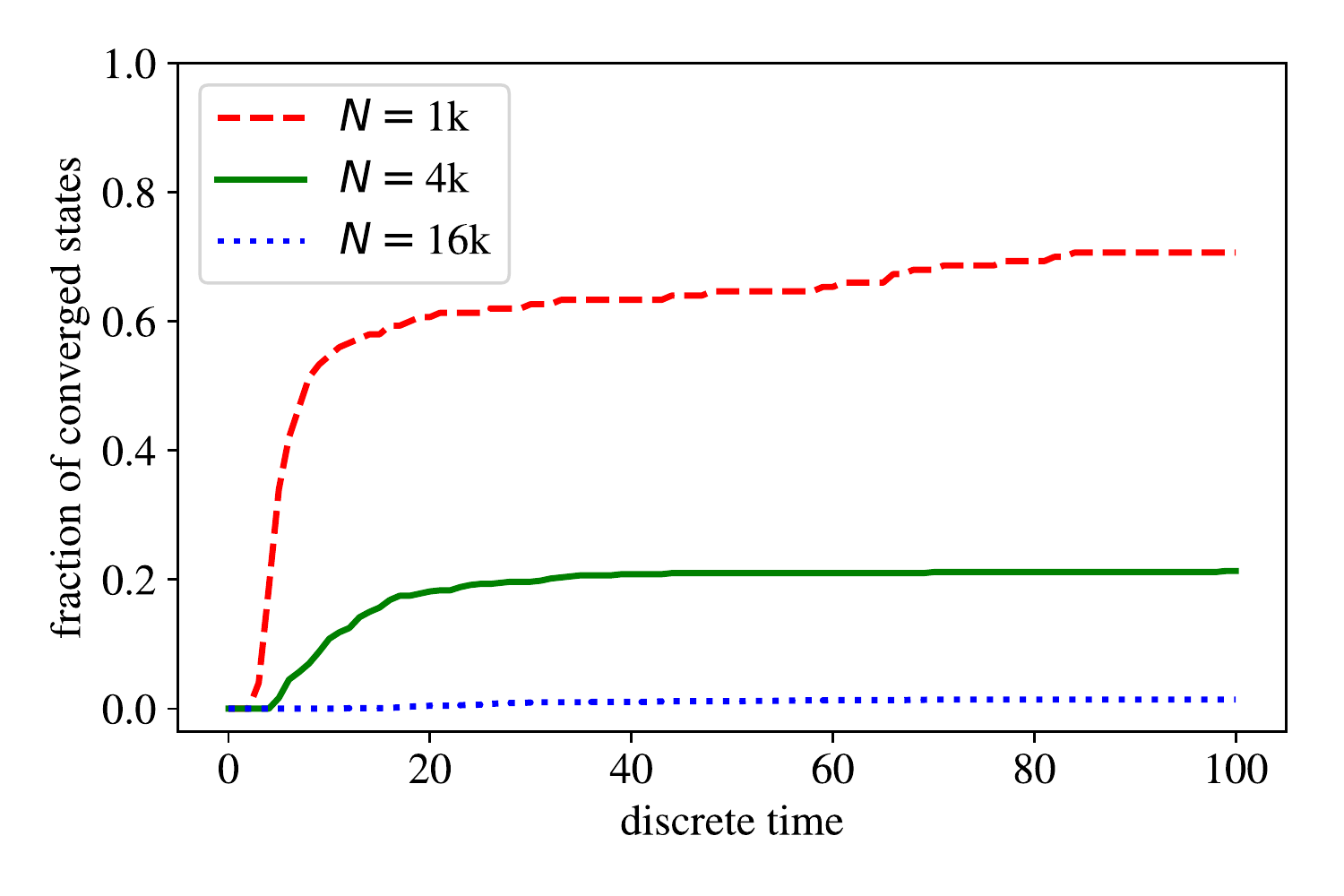}%
  }

\end{subfloatrow}

\begin{subfloatrow}[1]
\ffigbox[.99\textwidth]
  {
    \caption{dot-product evolution with $\alpha=0.13<\alpha_c$, $p=\alpha N \in \{130,520,2080\}$}
    \label{subfig:Lam0Dotp0}
  }
  {
\includegraphics[width=\linewidth]{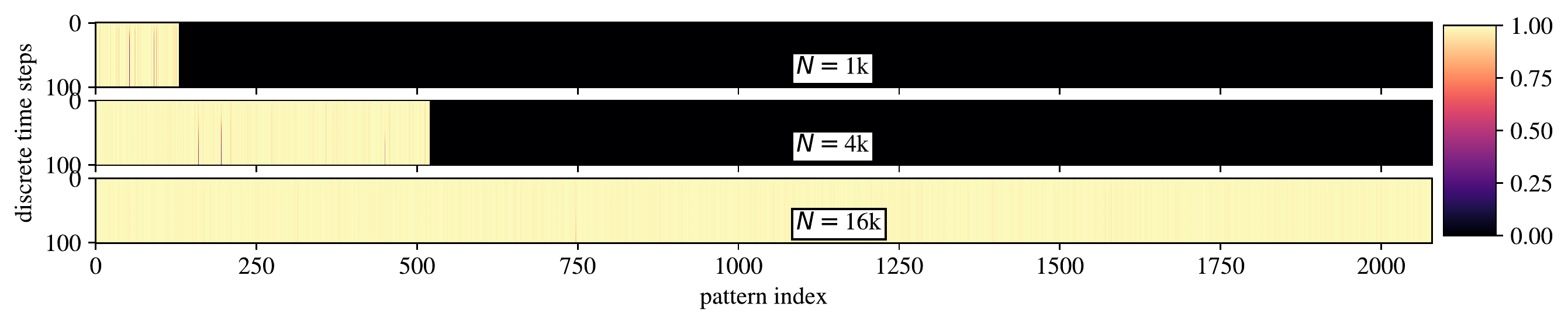}%
  }

\end{subfloatrow}

\begin{subfloatrow}[1]
\ffigbox[.99\textwidth]
  {
    \caption{dot-product evolution with $\alpha=0.15>\alpha_c$, $p=\alpha N \in \{150,600,2400\}$}
    \label{subfig:Lam0Dotp1}
  }
  {
\includegraphics[width=\linewidth]{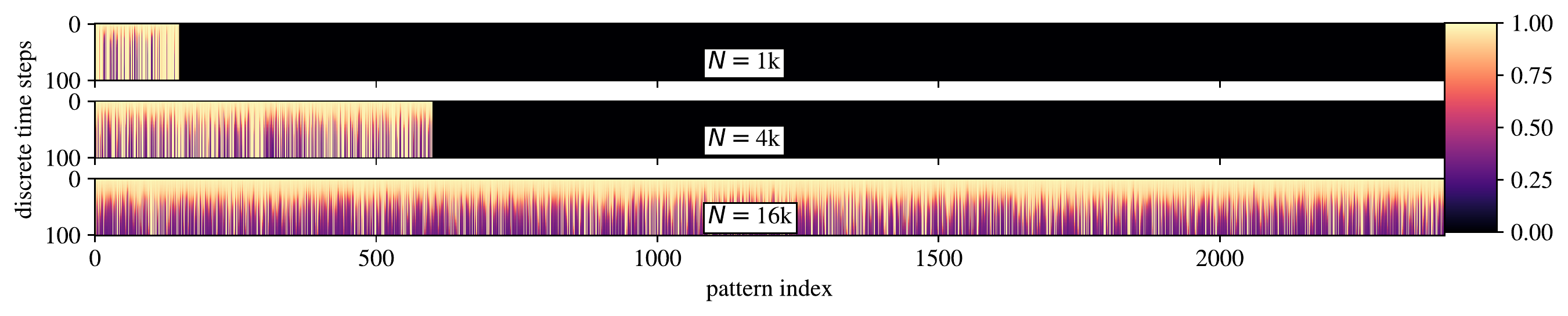}%
  }

\end{subfloatrow}

}
{
    \caption{Dynamics of the Hopfield network with zero self-coupling terms ($\lambda=0$), $N \in \{1k,4k,16k\}$, and initial overlap $m_0=0.9$. In (a),(c), and (e) we show dynamics for $\alpha=0.13$. In (b),(d), and (f) we show $\alpha=0.15$. We plot the probability distribution of overlaps $P(m_f)$ with mean $\mu(m_f)$ after 100 time steps in (a) and (b). Below criticality, the overlap at the $m=1$ weight increases with $N$, suggesting the memory loading is below criticality $\alpha_c$. In (b), the loading at $m=1$ decreases with $N$, implying that $\alpha > \alpha_c$. In (c) and (d) we observe that with increasing $N$ and synchronous update, patterns do not converge to a steady state and instead fluctuate in 2-cycles. In (e) and (f) we show how the overlap $m$ (represented by the colorbar) evolves over time.}
    \label{fig:Lam0Dyn}
}
\end{figure}

\begin{figure}[t]

\setlength\fboxsep{0pt}\setlength\fboxrule{0.75pt}
\ffigbox[\textwidth]
{
\begin{subfloatrow}[2]
\ffigbox[.45\textwidth]
  {
    \caption{$P(m)$ histogram with $\alpha=0.21<\alpha_c$}
    \label{subfig:Lam1Pm0}
  }
  {
\includegraphics[width=\linewidth]{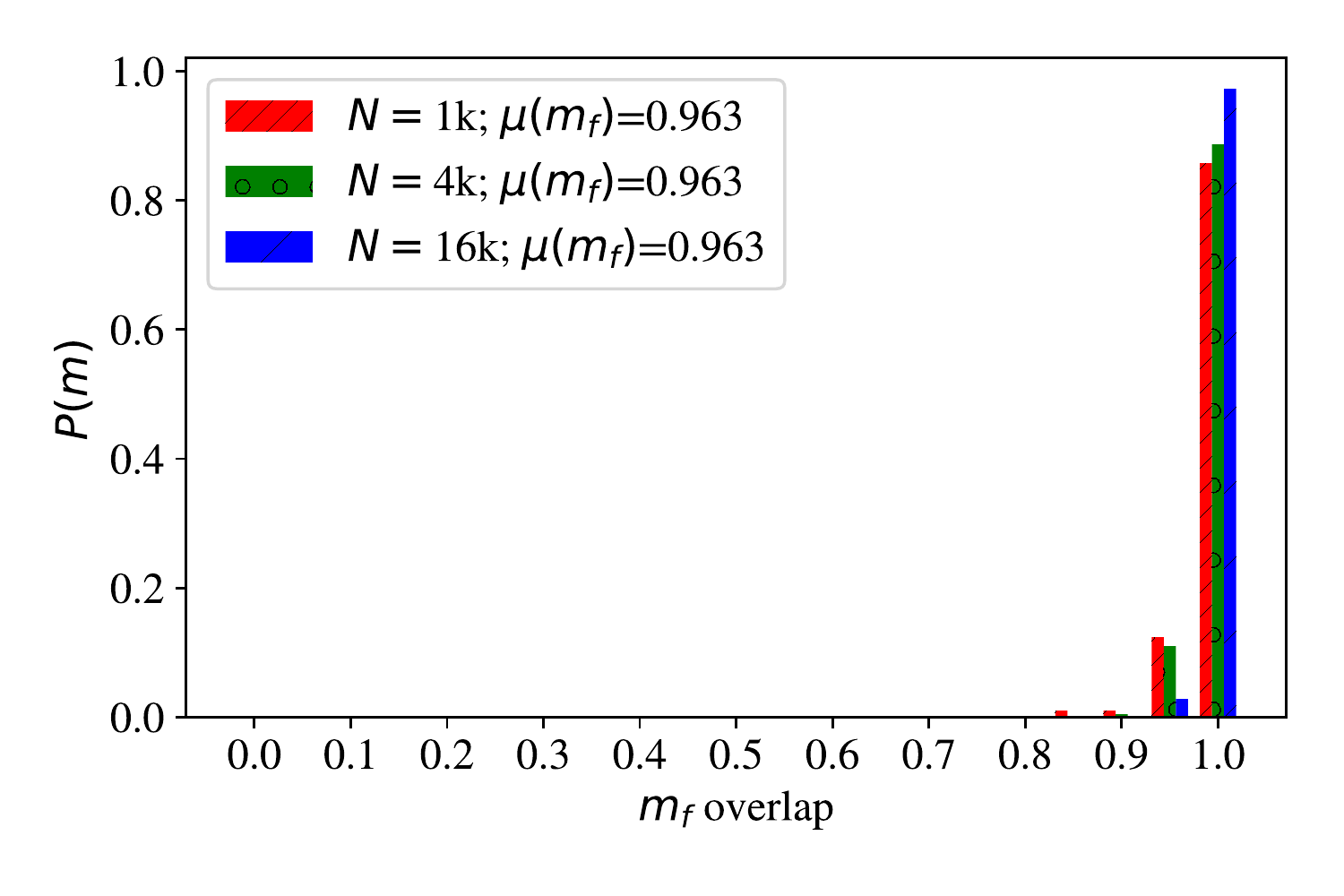}%
  }
\ffigbox[.45\textwidth]
  {
    \caption{$P(m)$ histogram with $\alpha=0.24>\alpha_c$}
    \label{subfig:Lam1Pm1}
  }
  {
    \includegraphics[width=\linewidth]{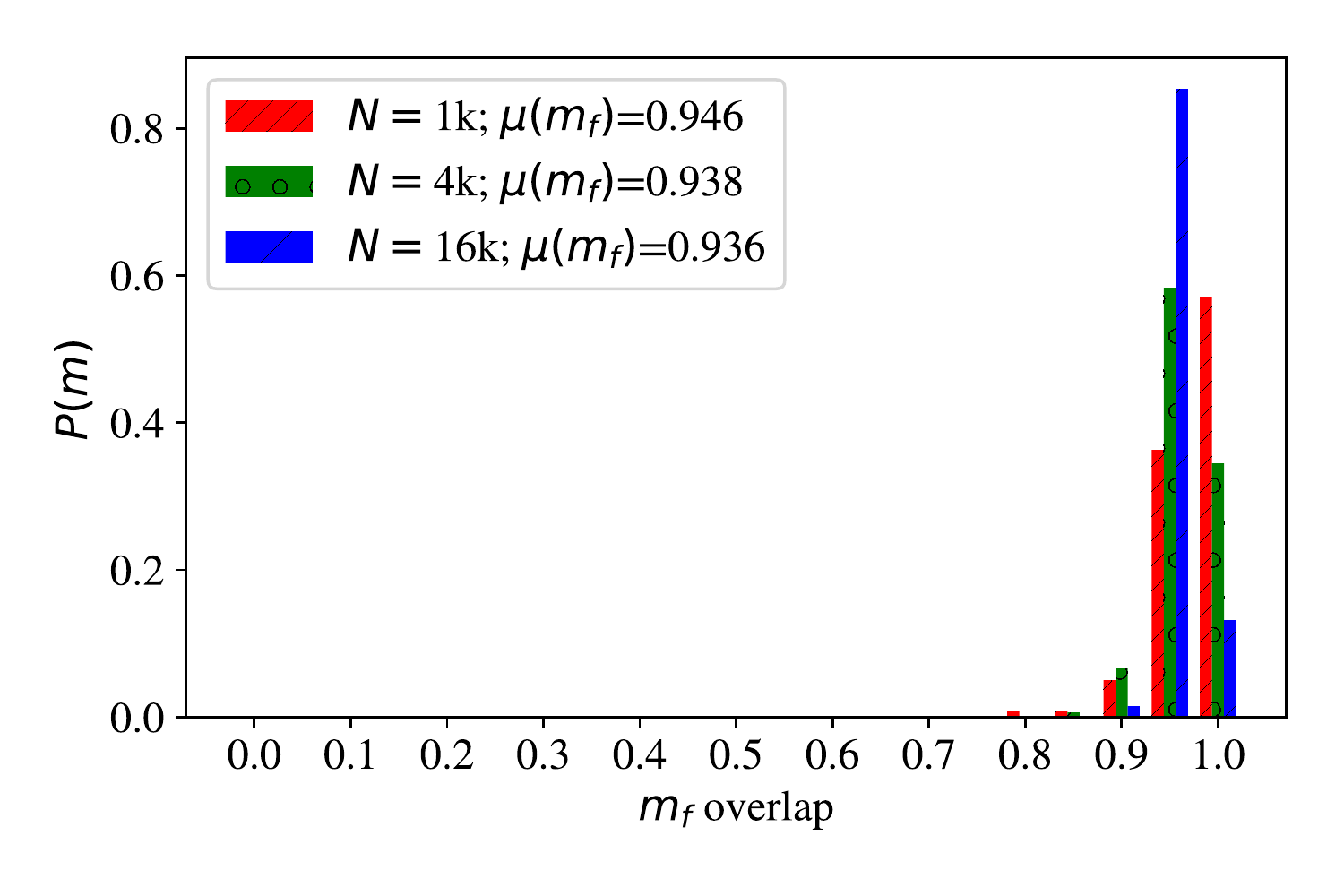}%
  }

\end{subfloatrow}

\begin{subfloatrow}[2]
\ffigbox[.45\textwidth]
  {
    \caption{pattern convergence with $\alpha=0.21<\alpha_c$}
    \label{subfig:Lam1Frac0}
  }
  {
\includegraphics[width=\linewidth]{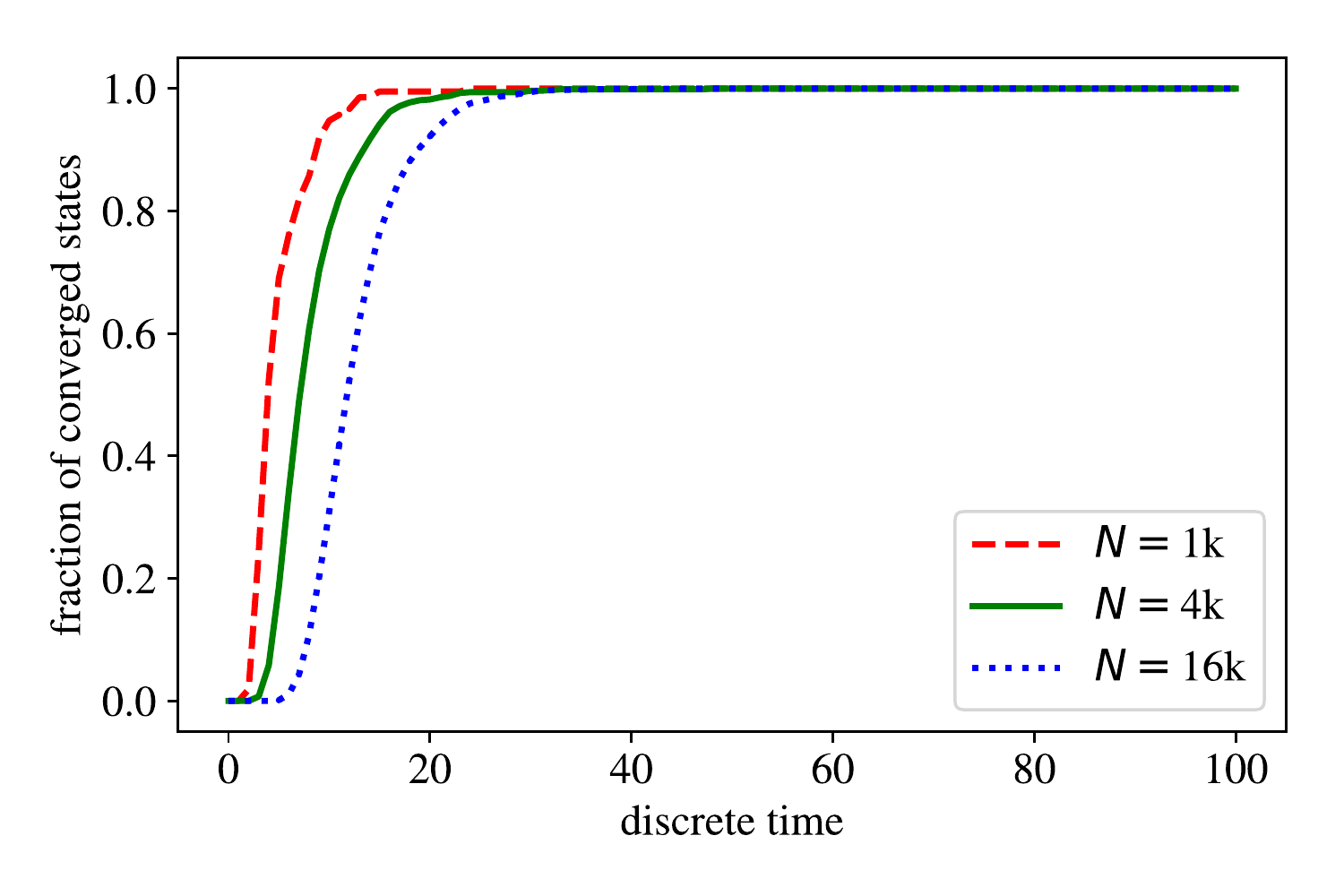}%
  }
\ffigbox[.45\textwidth]
  {
    \caption{pattern convergence with $\alpha=0.24>\alpha_c$}
    \label{subfig:Lam1Frac1}
  }
  {
    \includegraphics[width=\linewidth]{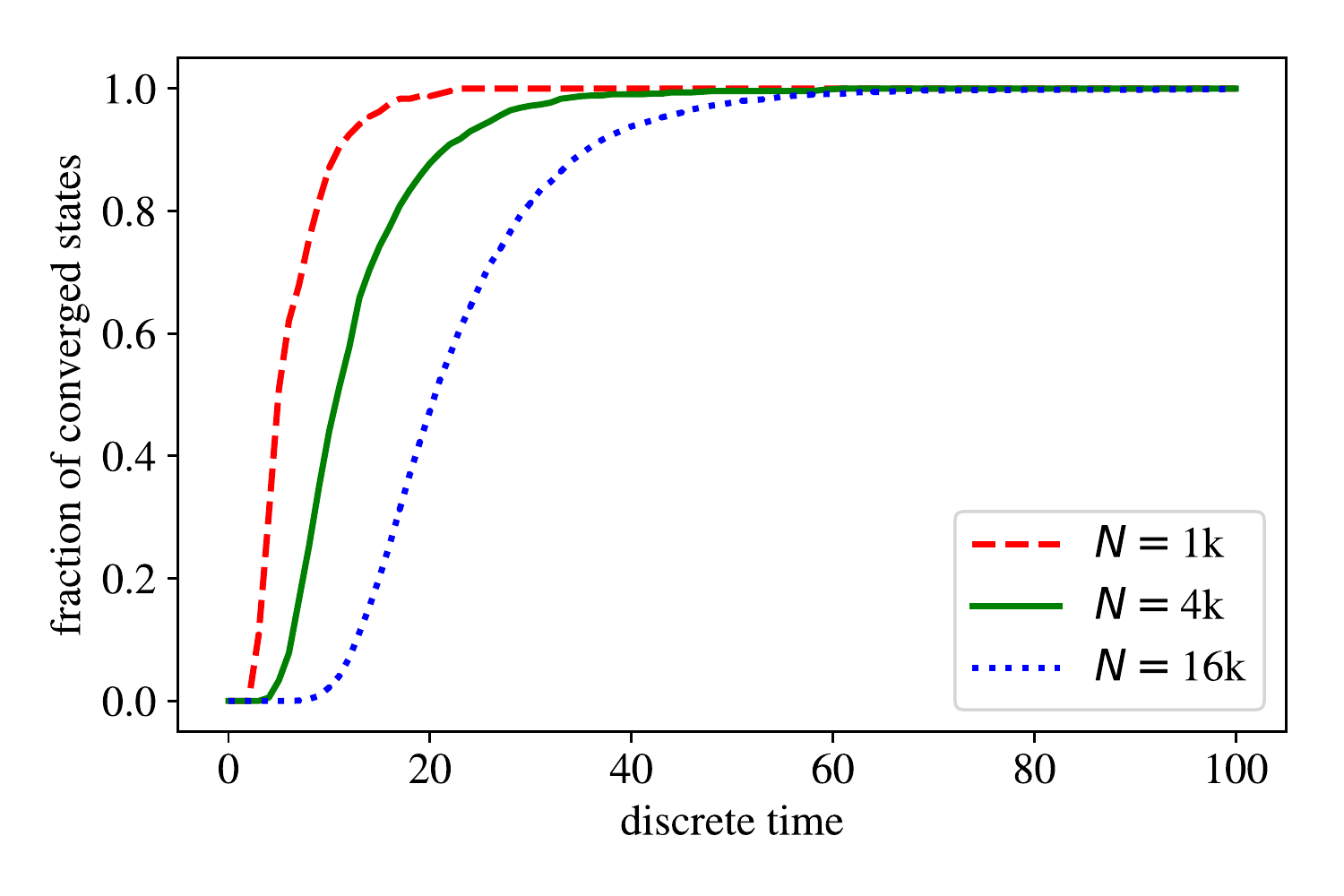}%
  }

\end{subfloatrow}

\begin{subfloatrow}[1]
\ffigbox[.99\textwidth]
  {
    \caption{dot-product evolution with $\alpha=0.21<\alpha_c$, $p=\alpha N \in \{210,840,3360\}$}
    \label{subfig:Lam1Dotp0}
  }
  {
\includegraphics[width=\linewidth]{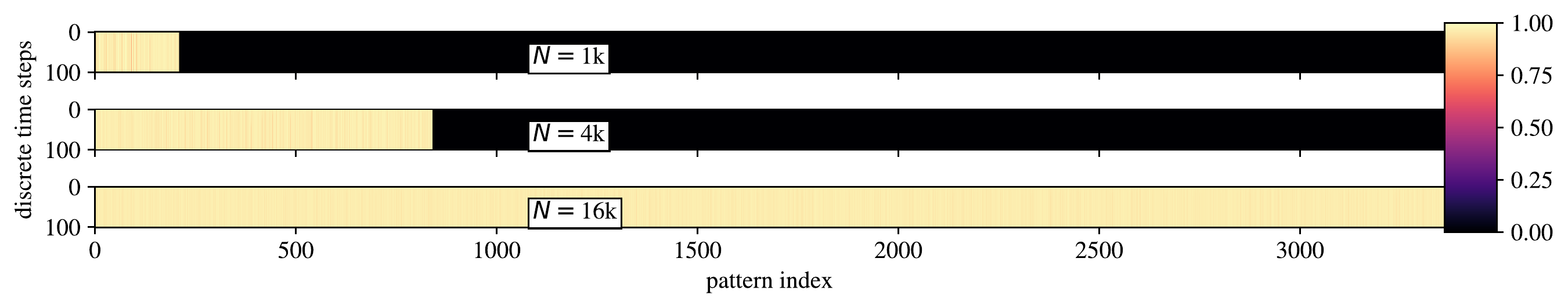}%
  }

\end{subfloatrow}

\begin{subfloatrow}[1]
\ffigbox[.99\textwidth]
  {
    \caption{dot-product evolution with $\alpha=0.24>\alpha_c$, $p=\alpha N \in \{240,960,3840\}$}
    \label{subfig:Lam1Dotp1}
  }
  {
\includegraphics[width=\linewidth]{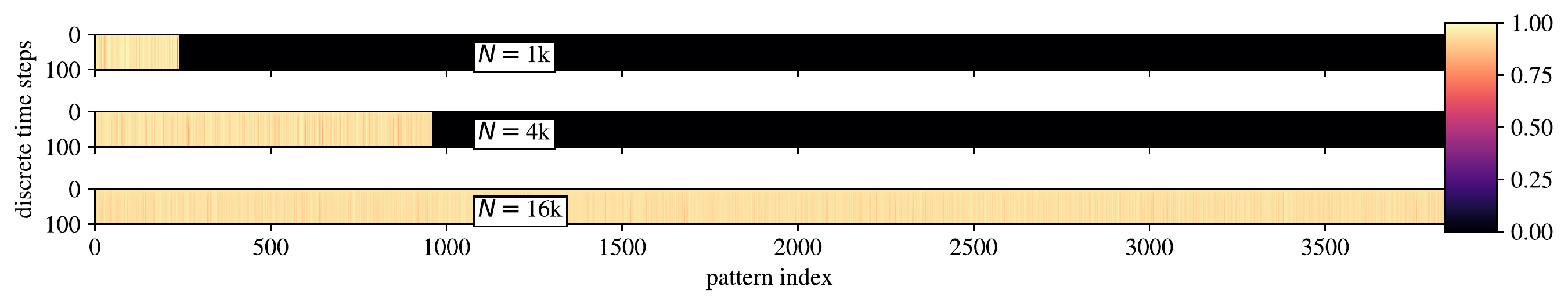}%
  }

\end{subfloatrow}
}
{
    \caption{Dynamics of the Hopfield network with self-coupling ($\lambda=1$), $N \in \{1k,4k,16k\}$, and initial overlap $m_0=0.9$. In (a),(c), and (e) we show dynamics for $\alpha=0.21$. In (b),(d), and (f) we show $\alpha=0.24$. We plot the probability distribution of overlaps $P(m_f)$ with mean $\mu(m_f)$ after 100 time steps in (a) and (b). Below criticality, the overlap at the $m=1$ weight increases with $N$, suggesting the memory loading is below criticality $\alpha_c$. In (b), the loading at $m=1$ decreases with $N$, implying that $\alpha > \alpha_c$. In (c) and (d) we observe that with synchronous update, all patterns converge but the rate slows with increasing $N$. In (e) and (f) we show how the overlap $m$ evolves over time.}
    \label{fig:Lam1Dyn}
}
\end{figure}

We now execute large-scale simulations and plot the results in figs.~\ref{fig:Lam0Dyn} and \ref{fig:Lam1Dyn}. In fig.~\ref{fig:Lam0Dyn}, we visualize the network with zero self-coupling terms ($\lambda=0$), $\alpha \in \{0.13,0.15\}$, $N \in \{1k,4k,16k\}$, and initial overlap $m_0=0.9$. To study network recall quality, we can plot the probability distribution of overlaps $P(m_f)$, shown in figs.~\ref{subfig:Lam0Pm0} and \ref{subfig:Lam0Pm1}. Below criticality, the weight at $m=1$ increases with increasing $N$. Above criticality, the weight at $m=1$ decreases with $N$ and we instead observe a two-peak structure with weight emerging near $m=0.35$. We conclude that for $\lambda=0$, $0.13 < \alpha_c < 0.15$.  In figs.~\ref{subfig:Lam0Frac0} and \ref{subfig:Lam0Frac1} we plot how the fraction of converged states evolves over time. Near criticality, a vanishingly small number of states truly converge and states instead relax into $2$-cycles, a known result accompanying synchronous update. We leave it as an exercise for the reader to show that with asynchronous update, the 2-cycle behavior is eliminated while the memory capacity remains the same. 

In addition to asynchronous update, there are numerous avenues to further investigate the Hopfield network, such as modifying the synaptic matrix (with disorder, non-local weights, or dilution), asymmetric neural updating rules, and even using the Hopfield network to solve problems in another domain. The extension we show for this paper is probing the effects of non-zero self-action ($\lambda \ne 0$). We do so in line 90 of the code. In fig.~\ref{fig:Lam1Dyn}, we observe that $\lambda=1$ produces useful recall as high as $\alpha=0.21$, and its performance degrades more gracefully in response to loading exceeding criticality ($\alpha=0.24$, figs.~\ref{subfig:Lam1Pm0} and \ref{subfig:Lam1Pm1}). However, with the introduction of self-coupling, the convergence time appears to increase (figs.~\ref{subfig:Lam1Frac0} and \ref{subfig:Lam1Frac1}).~\cite{tsuboshita2010statistical}

\FloatBarrier
\captionsetup{justification=centerlast}
\floatsetup[subfloat]{floatrowsep=myfill}

\FloatBarrier

\captionsetup{justification=centerlast}
\floatsetup[subfloat]{floatrowsep=myfill}

\FloatBarrier

\section{Suggested Exercises} 
\begin{enumerate}
  \item Implement asynchronous update by introducing a parameter $0<k<1$ such that at every time step, $kN$ neurons are updated. When $k=1/N$, this is purely asynchronous update. When $k=1$, we have purely synchronous update. For all other $k$, we have hybrid updating, which enjoys the massive parallelism inherent in synchronous update, while avoiding 2-cycles. 
  \item In the large-$N$ limit, the maximal number of memories stored such that all are recalled perfectly\cite{mbf} is $p<N/4log(N)$. Derive this result with theory using a signal-to-noise analysis\cite{mbf,baryam,greenBook} and test it with simulation.
  \item Stochastic noise is typically implemented by using a probabilistic update rule~\cite{mbf,dja1} that modifies the time evolution of each neuron as follows:
\begin{eqnarray}
h_i & = & \sum_{j\neq i}^{N} J_{ij} S_j(t)\\
P_{S_i}(+1;t+1) & = &  \left(1+\tanh\left(\beta h_i \right)\right)/2\\
P_{S_i}(-1;t+1) & = & 1 - P_{S_i}(+1;t+1)
\label{eq:TempEqn}
\end{eqnarray}
where $\beta = 1/k_BT$ tunes the strength of noise, $T$ is the absolute temperature, and $k_B$ is the Boltzmann constant. $\beta = 0$ encodes totally random dynamics, and $T=0$ encodes the usual deterministic update. Produce a $T-\alpha$ phase diagram to show the effect of these two parameters on $m_f$ for varying $m_0$ and large $N$. 
\end{enumerate}
\setstretch{2}
\section{Conclusion}

In this tutorial, we have suggested the use of cloud computing, GPUs, and deep learning frameworks to accelerate large-scale physical simulations and make high-performance computing accessible to students and researchers. We demonstrated this by performing a simulation of the Hopfield network with large system size ($N=32K$) and realized a GPU acceleration exceeding 50$\times$ as compared with CPU-only simulations -- using only free cloud resources. Our hope is that the rapid pace of development in the computer science discipline can enable physicists to work faster, and help educators remove barriers between their students and participation in research. We encourage the reader to modify the example code and implement their own physical simulations. 

\appendix*   

\section{Plotting code}

To aid student learning, we show the reader the source code to reproduce the fraction of converged states figures (\ref{subfig:Lam0Frac0}, \ref{subfig:Lam0Frac1}, \ref{subfig:Lam1Frac0}, \ref{subfig:Lam1Frac1}). The following code can be appended to the simulation code for instant viewing of figures in the browser. We use the Matplotlib package for creating publication-quality figures, and note that line 152 shows how to download any file (here a pdf image) from the cloud machine onto a personal computer. Implementations of the remaining figures look similar and we encourage students to reproduce them.

All source code in this paper can be found in this pre-populated Colab notebook: \url{https://colab.research.google.com/drive/1bS9V5GDzfeKe3Pu_yza8t66KjUNpMcFM}

\StartLineAt{128}
\singlespacing
\begin{lstlisting}[label=FracConv,caption=Plot fraction of converged states]
import matplotlib, matplotlib.pyplot as plt
from google.colab import files
font = {'family' : 'STIXGeneral',
        'weight' : 'normal',
        'size'   : 14}
matplotlib.rc('font', **font)
fig = plt.figure()
fig.set_size_inches(6, 4)
ax = plt.axes()
xticks = np.arange(0,max_steps+1,1)
ax.plot(xticks,num_conv_lists[0].numpy()[0,:],'--r',label=r'$N=$'+str(int(N_list[0]/1000))+'k',linewidth=2);
ax.plot(xticks,num_conv_lists[1].numpy()[0,:],'-g',label=r'$N=$'+str(int(N_list[1]/1000))+'k',linewidth=2);
ax.plot(xticks,num_conv_lists[2].numpy()[0,:],':b',label=r'$N=$'+str(int(N_list[2]/1000))+'k',linewidth=2);
plt.xlabel(r'discrete time')
plt.ylabel(r'fraction of converged states')
leg = plt.legend()
yticks = np.arange(0,1.001,0.2)
plt.yticks(yticks,[str(y)[0:3] for y in yticks])
leg_lines = leg.get_lines()
plt.setp(leg_lines, linewidth=2)
plt.tight_layout()
plt.show()
fname='frac_lambda0_alpha014.pdf'
fig.savefig(fname, dpi=300)
files.download(fname)
\end{lstlisting}
\setstretch{2}
\ContinueLineNumber
\begin{acknowledgments}

The author thanks Prof. Yogesh N. Joglekar for helpful conversations, comments, and revisions. This work was supported by the DJ Angus Foundation Summer Research Program and NSF grant no. DMR-1054020.   

\end{acknowledgments}

\bibliography{bib}

\begin{thebibliography}{31}%
\makeatletter
\providecommand \@ifxundefined [1]{%
 \@ifx{#1\undefined}
}%
\providecommand \@ifnum [1]{%
 \ifnum #1\expandafter \@firstoftwo
 \else \expandafter \@secondoftwo
 \fi
}%
\providecommand \@ifx [1]{%
 \ifx #1\expandafter \@firstoftwo
 \else \expandafter \@secondoftwo
 \fi
}%
\providecommand \natexlab [1]{#1}%
\providecommand \enquote  [1]{``#1''}%
\providecommand \bibnamefont  [1]{#1}%
\providecommand \bibfnamefont [1]{#1}%
\providecommand \citenamefont [1]{#1}%
\providecommand \href@noop [0]{\@secondoftwo}%
\providecommand \href [0]{\begingroup \@sanitize@url \@href}%
\providecommand \@href[1]{\@@startlink{#1}\@@href}%
\providecommand \@@href[1]{\endgroup#1\@@endlink}%
\providecommand \@sanitize@url [0]{\catcode `\\12\catcode `\$12\catcode
  `\&12\catcode `\#12\catcode `\^12\catcode `\_12\catcode `\%12\relax}%
\providecommand \@@startlink[1]{}%
\providecommand \@@endlink[0]{}%
\providecommand \url  [0]{\begingroup\@sanitize@url \@url }%
\providecommand \@url [1]{\endgroup\@href {#1}{\urlprefix }}%
\providecommand \urlprefix  [0]{URL }%
\providecommand \Eprint [0]{\href }%
\providecommand \doibase [0]{http://dx.doi.org/}%
\providecommand \selectlanguage [0]{\@gobble}%
\providecommand \bibinfo  [0]{\@secondoftwo}%
\providecommand \bibfield  [0]{\@secondoftwo}%
\providecommand \translation [1]{[#1]}%
\providecommand \BibitemOpen [0]{}%
\providecommand \bibitemStop [0]{}%
\providecommand \bibitemNoStop [0]{.\EOS\space}%
\providecommand \EOS [0]{\spacefactor3000\relax}%
\providecommand \BibitemShut  [1]{\csname bibitem#1\endcsname}%
\let\auto@bib@innerbib\@empty
\bibitem [{\citenamefont {Caballero}\ and\ \citenamefont
  {Pollock}(2014)}]{doi:10.1119/1.4837437}%
  \BibitemOpen
  \bibfield  {author} {\bibinfo {author} {\bibfnamefont {M.~D.}\ \bibnamefont
  {Caballero}}\ and\ \bibinfo {author} {\bibfnamefont {S.~J.}\ \bibnamefont
  {Pollock}},\ }\href {\doibase 10.1119/1.4837437} {\bibfield  {journal}
  {\bibinfo  {journal} {American Journal of Physics}\ }\textbf {\bibinfo
  {volume} {82}},\ \bibinfo {pages} {231} (\bibinfo {year} {2014})},\ \Eprint
  {http://arxiv.org/abs/https://doi.org/10.1119/1.4837437}
  {https://doi.org/10.1119/1.4837437} \BibitemShut {NoStop}%
\bibitem [{\citenamefont {Goodfellow}\ \emph {et~al.}(2016)\citenamefont
  {Goodfellow}, \citenamefont {Bengio},\ and\ \citenamefont
  {Courville}}]{Goodfellow:2016:DL:3086952}%
  \BibitemOpen
  \bibfield  {author} {\bibinfo {author} {\bibfnamefont {I.}~\bibnamefont
  {Goodfellow}}, \bibinfo {author} {\bibfnamefont {Y.}~\bibnamefont {Bengio}},
  \ and\ \bibinfo {author} {\bibfnamefont {A.}~\bibnamefont {Courville}},\
  }\href@noop {} {\emph {\bibinfo {title} {Deep Learning}}}\ (\bibinfo
  {publisher} {The MIT Press},\ \bibinfo {year} {2016})\BibitemShut {NoStop}%
\bibitem [{\citenamefont {Rumelhart}\ \emph {et~al.}(1986)\citenamefont
  {Rumelhart}, \citenamefont {Hinton},\ and\ \citenamefont
  {Williams}}]{Rumelhart:1986we}%
  \BibitemOpen
  \bibfield  {author} {\bibinfo {author} {\bibfnamefont {D.~E.}\ \bibnamefont
  {Rumelhart}}, \bibinfo {author} {\bibfnamefont {G.~E.}\ \bibnamefont
  {Hinton}}, \ and\ \bibinfo {author} {\bibfnamefont {R.~J.}\ \bibnamefont
  {Williams}},\ }\href {\doibase 10.1038/323533a0} {\bibfield  {journal}
  {\bibinfo  {journal} {Nature}\ }\textbf {\bibinfo {volume} {323}},\ \bibinfo
  {pages} {533} (\bibinfo {year} {1986})}\BibitemShut {NoStop}%
\bibitem [{\citenamefont {Coates}\ \emph {et~al.}(2013)\citenamefont {Coates},
  \citenamefont {Huval}, \citenamefont {Wang}, \citenamefont {Wu},
  \citenamefont {Catanzaro},\ and\ \citenamefont {Andrew}}]{pmlr-v28-coates13}%
  \BibitemOpen
  \bibfield  {author} {\bibinfo {author} {\bibfnamefont {A.}~\bibnamefont
  {Coates}}, \bibinfo {author} {\bibfnamefont {B.}~\bibnamefont {Huval}},
  \bibinfo {author} {\bibfnamefont {T.}~\bibnamefont {Wang}}, \bibinfo {author}
  {\bibfnamefont {D.}~\bibnamefont {Wu}}, \bibinfo {author} {\bibfnamefont
  {B.}~\bibnamefont {Catanzaro}}, \ and\ \bibinfo {author} {\bibfnamefont
  {N.}~\bibnamefont {Andrew}},\ }in\ \href
  {http://proceedings.mlr.press/v28/coates13.html} {\emph {\bibinfo {booktitle}
  {Proceedings of the 30th International Conference on Machine Learning}}},\
  \bibinfo {series} {Proceedings of Machine Learning Research}, Vol.~\bibinfo
  {volume} {28},\ \bibinfo {editor} {edited by\ \bibinfo {editor}
  {\bibfnamefont {S.}~\bibnamefont {Dasgupta}}\ and\ \bibinfo {editor}
  {\bibfnamefont {D.}~\bibnamefont {McAllester}}}\ (\bibinfo  {publisher}
  {PMLR},\ \bibinfo {address} {Atlanta, Georgia, USA},\ \bibinfo {year}
  {2013})\ pp.\ \bibinfo {pages} {1337--1345}\BibitemShut {NoStop}%
\bibitem [{\citenamefont {Akenine-Moller}\ \emph {et~al.}(2008)\citenamefont
  {Akenine-Moller}, \citenamefont {Haines},\ and\ \citenamefont
  {Hoffman}}]{Akenine-Moller:2008:RR:2829183}%
  \BibitemOpen
  \bibfield  {author} {\bibinfo {author} {\bibfnamefont {T.}~\bibnamefont
  {Akenine-Moller}}, \bibinfo {author} {\bibfnamefont {E.}~\bibnamefont
  {Haines}}, \ and\ \bibinfo {author} {\bibfnamefont {N.}~\bibnamefont
  {Hoffman}},\ }\href@noop {} {\emph {\bibinfo {title} {Real-Time
  Rendering}}},\ \bibinfo {edition} {3rd}\ ed.\ (\bibinfo  {publisher} {A. K.
  Peters, Ltd.},\ \bibinfo {address} {Natick, MA, USA},\ \bibinfo {year}
  {2008})\BibitemShut {NoStop}%
\bibitem [{\citenamefont {Kim}\ \emph {et~al.}(2019)\citenamefont {Kim},
  \citenamefont {Azevedo}, \citenamefont {Thuerey}, \citenamefont {Kim},
  \citenamefont {Gross},\ and\ \citenamefont
  {Solenthaler}}]{DBLP:journals/corr/abs-1806-02071}%
  \BibitemOpen
  \bibfield  {author} {\bibinfo {author} {\bibfnamefont {B.}~\bibnamefont
  {Kim}}, \bibinfo {author} {\bibfnamefont {V.~C.}\ \bibnamefont {Azevedo}},
  \bibinfo {author} {\bibfnamefont {N.}~\bibnamefont {Thuerey}}, \bibinfo
  {author} {\bibfnamefont {T.}~\bibnamefont {Kim}}, \bibinfo {author}
  {\bibfnamefont {M.}~\bibnamefont {Gross}}, \ and\ \bibinfo {author}
  {\bibfnamefont {B.}~\bibnamefont {Solenthaler}},\ }\href {\doibase
  10.1111/cgf.13619} {\bibfield  {journal} {\bibinfo  {journal} {Computer
  Graphics Forum}\ }\textbf {\bibinfo {volume} {38}},\ \bibinfo {pages} {59}
  (\bibinfo {year} {2019})},\ \Eprint
  {http://arxiv.org/abs/https://onlinelibrary.wiley.com/doi/pdf/10.1111/cgf.13619}
  {https://onlinelibrary.wiley.com/doi/pdf/10.1111/cgf.13619} \BibitemShut
  {NoStop}%
\bibitem [{\citenamefont {Kavan}\ \emph {et~al.}(2011)\citenamefont {Kavan},
  \citenamefont {Gerszewski}, \citenamefont {Bargteil},\ and\ \citenamefont
  {Sloan}}]{kavan2011physics}%
  \BibitemOpen
  \bibfield  {author} {\bibinfo {author} {\bibfnamefont {L.}~\bibnamefont
  {Kavan}}, \bibinfo {author} {\bibfnamefont {D.}~\bibnamefont {Gerszewski}},
  \bibinfo {author} {\bibfnamefont {A.~W.}\ \bibnamefont {Bargteil}}, \ and\
  \bibinfo {author} {\bibfnamefont {P.-P.}\ \bibnamefont {Sloan}},\ }in\
  \href@noop {} {\emph {\bibinfo {booktitle} {ACM Transactions on Graphics
  (TOG)}}},\ Vol.~\bibinfo {volume} {30}\ (\bibinfo {organization} {ACM},\
  \bibinfo {year} {2011})\ p.~\bibinfo {pages} {93}\BibitemShut {NoStop}%
\bibitem [{\citenamefont {Tariq}\ and\ \citenamefont
  {Bavoil}(2008)}]{Tariq2008RealTH}%
  \BibitemOpen
  \bibfield  {author} {\bibinfo {author} {\bibfnamefont {S.}~\bibnamefont
  {Tariq}}\ and\ \bibinfo {author} {\bibfnamefont {L.}~\bibnamefont {Bavoil}},\
  }in\ \href@noop {} {\emph {\bibinfo {booktitle} {SIGGRAPH '08}}}\ (\bibinfo
  {year} {2008})\BibitemShut {NoStop}%
\bibitem [{\citenamefont {Paszke}\ \emph {et~al.}(2019)\citenamefont {Paszke},
  \citenamefont {Gross}, \citenamefont {Massa}, \citenamefont {Lerer},
  \citenamefont {Bradbury}, \citenamefont {Chanan}, \citenamefont {Killeen},
  \citenamefont {Lin}, \citenamefont {Gimelshein}, \citenamefont {Antiga},
  \citenamefont {Desmaison}, \citenamefont {Kopf}, \citenamefont {Yang},
  \citenamefont {DeVito}, \citenamefont {Raison}, \citenamefont {Tejani},
  \citenamefont {Chilamkurthy}, \citenamefont {Steiner}, \citenamefont {Fang},
  \citenamefont {Bai},\ and\ \citenamefont {Chintala}}]{NIPS2019_9015}%
  \BibitemOpen
  \bibfield  {author} {\bibinfo {author} {\bibfnamefont {A.}~\bibnamefont
  {Paszke}}, \bibinfo {author} {\bibfnamefont {S.}~\bibnamefont {Gross}},
  \bibinfo {author} {\bibfnamefont {F.}~\bibnamefont {Massa}}, \bibinfo
  {author} {\bibfnamefont {A.}~\bibnamefont {Lerer}}, \bibinfo {author}
  {\bibfnamefont {J.}~\bibnamefont {Bradbury}}, \bibinfo {author}
  {\bibfnamefont {G.}~\bibnamefont {Chanan}}, \bibinfo {author} {\bibfnamefont
  {T.}~\bibnamefont {Killeen}}, \bibinfo {author} {\bibfnamefont
  {Z.}~\bibnamefont {Lin}}, \bibinfo {author} {\bibfnamefont {N.}~\bibnamefont
  {Gimelshein}}, \bibinfo {author} {\bibfnamefont {L.}~\bibnamefont {Antiga}},
  \bibinfo {author} {\bibfnamefont {A.}~\bibnamefont {Desmaison}}, \bibinfo
  {author} {\bibfnamefont {A.}~\bibnamefont {Kopf}}, \bibinfo {author}
  {\bibfnamefont {E.}~\bibnamefont {Yang}}, \bibinfo {author} {\bibfnamefont
  {Z.}~\bibnamefont {DeVito}}, \bibinfo {author} {\bibfnamefont
  {M.}~\bibnamefont {Raison}}, \bibinfo {author} {\bibfnamefont
  {A.}~\bibnamefont {Tejani}}, \bibinfo {author} {\bibfnamefont
  {S.}~\bibnamefont {Chilamkurthy}}, \bibinfo {author} {\bibfnamefont
  {B.}~\bibnamefont {Steiner}}, \bibinfo {author} {\bibfnamefont
  {L.}~\bibnamefont {Fang}}, \bibinfo {author} {\bibfnamefont {J.}~\bibnamefont
  {Bai}}, \ and\ \bibinfo {author} {\bibfnamefont {S.}~\bibnamefont
  {Chintala}},\ }in\ \href
  {http://papers.nips.cc/paper/9015-pytorch-an-imperative-style-high-performance-deep-learning-library.pdf}
  {\emph {\bibinfo {booktitle} {Advances in Neural Information Processing
  Systems 32}}},\ \bibinfo {editor} {edited by\ \bibinfo {editor}
  {\bibfnamefont {H.}~\bibnamefont {Wallach}}, \bibinfo {editor} {\bibfnamefont
  {H.}~\bibnamefont {Larochelle}}, \bibinfo {editor} {\bibfnamefont
  {A.}~\bibnamefont {Beygelzimer}}, \bibinfo {editor} {\bibfnamefont
  {F.}~\bibnamefont {d\textquotesingle Alch\'{e}-Buc}}, \bibinfo {editor}
  {\bibfnamefont {E.}~\bibnamefont {Fox}}, \ and\ \bibinfo {editor}
  {\bibfnamefont {R.}~\bibnamefont {Garnett}}}\ (\bibinfo  {publisher} {Curran
  Associates, Inc.},\ \bibinfo {year} {2019})\ pp.\ \bibinfo {pages}
  {8024--8035}\BibitemShut {NoStop}%
\bibitem [{\citenamefont {Abadi}\ \emph {et~al.}(2016)\citenamefont {Abadi},
  \citenamefont {Agarwal}, \citenamefont {Barham}, \citenamefont {Brevdo},
  \citenamefont {Chen}, \citenamefont {Citro}, \citenamefont {Corrado},
  \citenamefont {Davis}, \citenamefont {Dean}, \citenamefont {Devin} \emph
  {et~al.}}]{tensorflow2015-whitepaper}%
  \BibitemOpen
  \bibfield  {author} {\bibinfo {author} {\bibfnamefont {M.}~\bibnamefont
  {Abadi}}, \bibinfo {author} {\bibfnamefont {A.}~\bibnamefont {Agarwal}},
  \bibinfo {author} {\bibfnamefont {P.}~\bibnamefont {Barham}}, \bibinfo
  {author} {\bibfnamefont {E.}~\bibnamefont {Brevdo}}, \bibinfo {author}
  {\bibfnamefont {Z.}~\bibnamefont {Chen}}, \bibinfo {author} {\bibfnamefont
  {C.}~\bibnamefont {Citro}}, \bibinfo {author} {\bibfnamefont {G.~S.}\
  \bibnamefont {Corrado}}, \bibinfo {author} {\bibfnamefont {A.}~\bibnamefont
  {Davis}}, \bibinfo {author} {\bibfnamefont {J.}~\bibnamefont {Dean}},
  \bibinfo {author} {\bibfnamefont {M.}~\bibnamefont {Devin}},  \emph
  {et~al.},\ }\href@noop {} {\bibfield  {journal} {\bibinfo  {journal} {arXiv
  preprint arXiv:1603.04467}\ } (\bibinfo {year} {2016})}\BibitemShut {NoStop}%
\bibitem [{\citenamefont {{Carneiro}}\ \emph {et~al.}(2018)\citenamefont
  {{Carneiro}}, \citenamefont {{Medeiros Da Nóbrega}}, \citenamefont
  {{Nepomuceno}}, \citenamefont {{Bian}}, \citenamefont {{De Albuquerque}},\
  and\ \citenamefont {{Filho}}}]{8485684}%
  \BibitemOpen
  \bibfield  {author} {\bibinfo {author} {\bibfnamefont {T.}~\bibnamefont
  {{Carneiro}}}, \bibinfo {author} {\bibfnamefont {R.~V.}\ \bibnamefont
  {{Medeiros Da Nóbrega}}}, \bibinfo {author} {\bibfnamefont {T.}~\bibnamefont
  {{Nepomuceno}}}, \bibinfo {author} {\bibfnamefont {G.}~\bibnamefont
  {{Bian}}}, \bibinfo {author} {\bibfnamefont {V.~H.~C.}\ \bibnamefont {{De
  Albuquerque}}}, \ and\ \bibinfo {author} {\bibfnamefont {P.~P.~R.}\
  \bibnamefont {{Filho}}},\ }\href {\doibase 10.1109/ACCESS.2018.2874767}
  {\bibfield  {journal} {\bibinfo  {journal} {IEEE Access}\ }\textbf {\bibinfo
  {volume} {6}},\ \bibinfo {pages} {61677} (\bibinfo {year}
  {2018})}\BibitemShut {NoStop}%
\bibitem [{\citenamefont {Okuta}\ \emph {et~al.}(2017)\citenamefont {Okuta},
  \citenamefont {Unno}, \citenamefont {Nishino}, \citenamefont {Hido},\ and\
  \citenamefont {Loomis}}]{okuta2017cupy}%
  \BibitemOpen
  \bibfield  {author} {\bibinfo {author} {\bibfnamefont {R.}~\bibnamefont
  {Okuta}}, \bibinfo {author} {\bibfnamefont {Y.}~\bibnamefont {Unno}},
  \bibinfo {author} {\bibfnamefont {D.}~\bibnamefont {Nishino}}, \bibinfo
  {author} {\bibfnamefont {S.}~\bibnamefont {Hido}}, \ and\ \bibinfo {author}
  {\bibfnamefont {C.}~\bibnamefont {Loomis}},\ }in\ \href@noop {} {\emph
  {\bibinfo {booktitle} {of Workshop on Machine Learning Systems (LearningSys)
  in The Thirty-first Annual Conference on Neural Information Processing
  Systems (NIPS)}}}\ (\bibinfo {year} {2017})\BibitemShut {NoStop}%
\bibitem [{\citenamefont {{Jouppi}}\ \emph {et~al.}(2018)\citenamefont
  {{Jouppi}}, \citenamefont {{Young}}, \citenamefont {{Patil}},\ and\
  \citenamefont {{Patterson}}}]{8358031}%
  \BibitemOpen
  \bibfield  {author} {\bibinfo {author} {\bibfnamefont {N.}~\bibnamefont
  {{Jouppi}}}, \bibinfo {author} {\bibfnamefont {C.}~\bibnamefont {{Young}}},
  \bibinfo {author} {\bibfnamefont {N.}~\bibnamefont {{Patil}}}, \ and\
  \bibinfo {author} {\bibfnamefont {D.}~\bibnamefont {{Patterson}}},\ }\href
  {\doibase 10.1109/MM.2018.032271057} {\bibfield  {journal} {\bibinfo
  {journal} {IEEE Micro}\ }\textbf {\bibinfo {volume} {38}},\ \bibinfo {pages}
  {10} (\bibinfo {year} {2018})}\BibitemShut {NoStop}%
\bibitem [{\citenamefont {Baldi}\ \emph {et~al.}(2014)\citenamefont {Baldi},
  \citenamefont {Sadowski},\ and\ \citenamefont
  {Whiteson}}]{baldi2014searching}%
  \BibitemOpen
  \bibfield  {author} {\bibinfo {author} {\bibfnamefont {P.}~\bibnamefont
  {Baldi}}, \bibinfo {author} {\bibfnamefont {P.}~\bibnamefont {Sadowski}}, \
  and\ \bibinfo {author} {\bibfnamefont {D.}~\bibnamefont {Whiteson}},\
  }\href@noop {} {\bibfield  {journal} {\bibinfo  {journal} {Nature
  communications}\ }\textbf {\bibinfo {volume} {5}},\ \bibinfo {pages} {4308}
  (\bibinfo {year} {2014})}\BibitemShut {NoStop}%
\bibitem [{\citenamefont {Mehta}\ \emph {et~al.}(2019)\citenamefont {Mehta},
  \citenamefont {Bukov}, \citenamefont {Wang}, \citenamefont {Day},
  \citenamefont {Richardson}, \citenamefont {Fisher},\ and\ \citenamefont
  {Schwab}}]{mehta2019high}%
  \BibitemOpen
  \bibfield  {author} {\bibinfo {author} {\bibfnamefont {P.}~\bibnamefont
  {Mehta}}, \bibinfo {author} {\bibfnamefont {M.}~\bibnamefont {Bukov}},
  \bibinfo {author} {\bibfnamefont {C.-H.}\ \bibnamefont {Wang}}, \bibinfo
  {author} {\bibfnamefont {A.~G.}\ \bibnamefont {Day}}, \bibinfo {author}
  {\bibfnamefont {C.}~\bibnamefont {Richardson}}, \bibinfo {author}
  {\bibfnamefont {C.~K.}\ \bibnamefont {Fisher}}, \ and\ \bibinfo {author}
  {\bibfnamefont {D.~J.}\ \bibnamefont {Schwab}},\ }\href {\doibase
  https://doi.org/10.1016/j.physrep.2019.03.001} {\bibfield  {journal}
  {\bibinfo  {journal} {Physics Reports}\ }\textbf {\bibinfo {volume} {810}},\
  \bibinfo {pages} {1 } (\bibinfo {year} {2019})},\ \bibinfo {note} {a
  high-bias, low-variance introduction to Machine Learning for
  physicists}\BibitemShut {NoStop}%
\bibitem [{\citenamefont {Schütt}\ \emph {et~al.}(2018)\citenamefont {Schütt},
  \citenamefont {Kessel}, \citenamefont {Gastegger}, \citenamefont {Nicoli},
  \citenamefont {Tkatchenko},\ and\ \citenamefont
  {Müller}}]{schütt2018schnetpack}%
  \BibitemOpen
  \bibfield  {author} {\bibinfo {author} {\bibfnamefont {K.}~\bibnamefont
  {Schütt}}, \bibinfo {author} {\bibfnamefont {P.}~\bibnamefont {Kessel}},
  \bibinfo {author} {\bibfnamefont {M.}~\bibnamefont {Gastegger}}, \bibinfo
  {author} {\bibfnamefont {K.}~\bibnamefont {Nicoli}}, \bibinfo {author}
  {\bibfnamefont {A.}~\bibnamefont {Tkatchenko}}, \ and\ \bibinfo {author}
  {\bibfnamefont {K.-R.}\ \bibnamefont {Müller}},\ }\href@noop {} {\bibfield
  {journal} {\bibinfo  {journal} {Journal of chemical theory and computation}\
  }\textbf {\bibinfo {volume} {15}},\ \bibinfo {pages} {448} (\bibinfo {year}
  {2018})}\BibitemShut {NoStop}%
\bibitem [{\citenamefont {Hopfield}(1982)}]{hop}%
  \BibitemOpen
  \bibfield  {author} {\bibinfo {author} {\bibfnamefont {J.~J.}\ \bibnamefont
  {Hopfield}},\ }\href {\doibase 10.1073/pnas.79.8.2554} {\bibfield  {journal}
  {\bibinfo  {journal} {Proceedings of the National Academy of Sciences}\
  }\textbf {\bibinfo {volume} {79}},\ \bibinfo {pages} {2554} (\bibinfo {year}
  {1982})}\BibitemShut {NoStop}%
\bibitem [{\citenamefont {Amit}\ \emph {et~al.}(1987)\citenamefont {Amit},
  \citenamefont {Gutfreund},\ and\ \citenamefont {Sompolinsky}}]{dja1}%
  \BibitemOpen
  \bibfield  {author} {\bibinfo {author} {\bibfnamefont {D.~J.}\ \bibnamefont
  {Amit}}, \bibinfo {author} {\bibfnamefont {H.}~\bibnamefont {Gutfreund}}, \
  and\ \bibinfo {author} {\bibfnamefont {H.}~\bibnamefont {Sompolinsky}},\
  }\href {\doibase 10.1016/0003-4916(87)90092-3} {\bibfield  {journal}
  {\bibinfo  {journal} {Annals of Physics}\ }\textbf {\bibinfo {volume}
  {173}},\ \bibinfo {pages} {30} (\bibinfo {year} {1987})}\BibitemShut
  {NoStop}%
\bibitem [{\citenamefont {Bar-Yam}(1997)}]{baryam}%
  \BibitemOpen
  \bibfield  {author} {\bibinfo {author} {\bibfnamefont {Y.}~\bibnamefont
  {Bar-Yam}},\ }\href@noop {} {\emph {\bibinfo {title} {Dynamics of complex
  systems}}}\ (\bibinfo  {publisher} {Addison-Wesley},\ \bibinfo {year}
  {1997})\BibitemShut {NoStop}%
\bibitem [{\citenamefont {Li}\ \emph {et~al.}(2005)\citenamefont {Li},
  \citenamefont {Tang}, \citenamefont {Xia},\ and\ \citenamefont
  {Wang}}]{crct}%
  \BibitemOpen
  \bibfield  {author} {\bibinfo {author} {\bibfnamefont {Y.}~\bibnamefont
  {Li}}, \bibinfo {author} {\bibfnamefont {Z.}~\bibnamefont {Tang}}, \bibinfo
  {author} {\bibfnamefont {G.}~\bibnamefont {Xia}}, \ and\ \bibinfo {author}
  {\bibfnamefont {R.}~\bibnamefont {Wang}},\ }\href {\doibase
  10.1109/TCSI.2004.838146} {\bibfield  {journal} {\bibinfo  {journal}
  {Circuits and Systems I: Regular Papers, IEEE Transactions on}\ }\textbf
  {\bibinfo {volume} {52}},\ \bibinfo {pages} {200} (\bibinfo {year}
  {2005})}\BibitemShut {NoStop}%
\bibitem [{\citenamefont {Li}\ \emph {et~al.}(2004)\citenamefont {Li},
  \citenamefont {Tang}, \citenamefont {Xia}, \citenamefont {Wang},\ and\
  \citenamefont {Xu}}]{tsp}%
  \BibitemOpen
  \bibfield  {author} {\bibinfo {author} {\bibfnamefont {Y.}~\bibnamefont
  {Li}}, \bibinfo {author} {\bibfnamefont {Z.}~\bibnamefont {Tang}}, \bibinfo
  {author} {\bibfnamefont {G.}~\bibnamefont {Xia}}, \bibinfo {author}
  {\bibfnamefont {R.~L.}\ \bibnamefont {Wang}}, \ and\ \bibinfo {author}
  {\bibfnamefont {X.}~\bibnamefont {Xu}},\ }in\ \href@noop {} {\emph {\bibinfo
  {booktitle} {SICE 2004 Annual Conference}}},\ Vol.~\bibinfo {volume} {2}\
  (\bibinfo {year} {2004})\ pp.\ \bibinfo {pages} {999--1004 vol.
  2}\BibitemShut {NoStop}%
\bibitem [{\citenamefont {Kazemi}\ \emph {et~al.}(2008)\citenamefont {Kazemi},
  \citenamefont {Akbarzadeh-T}, \citenamefont {Rahati},\ and\ \citenamefont
  {Rajabi}}]{imseg}%
  \BibitemOpen
  \bibfield  {author} {\bibinfo {author} {\bibfnamefont {F.}~\bibnamefont
  {Kazemi}}, \bibinfo {author} {\bibfnamefont {M.-R.}\ \bibnamefont
  {Akbarzadeh-T}}, \bibinfo {author} {\bibfnamefont {S.}~\bibnamefont
  {Rahati}}, \ and\ \bibinfo {author} {\bibfnamefont {H.}~\bibnamefont
  {Rajabi}},\ }in\ \href {\doibase 10.1109/CCECE.2008.4564866} {\emph {\bibinfo
  {booktitle} {Electrical and Computer Engineering, 2008. CCECE 2008. Canadian
  Conference on}}}\ (\bibinfo {year} {2008})\ pp.\ \bibinfo {pages}
  {001855--001860}\BibitemShut {NoStop}%
\bibitem [{\citenamefont {Widodo}\ \emph {et~al.}(2018)\citenamefont {Widodo},
  \citenamefont {Priambodo},\ and\ \citenamefont {Adhi}}]{Widodo_2018}%
  \BibitemOpen
  \bibfield  {author} {\bibinfo {author} {\bibfnamefont {W.}~\bibnamefont
  {Widodo}}, \bibinfo {author} {\bibfnamefont {R.~A.}\ \bibnamefont
  {Priambodo}}, \ and\ \bibinfo {author} {\bibfnamefont {B.~P.}\ \bibnamefont
  {Adhi}},\ }\href {\doibase 10.1088/1757-899x/434/1/012034} {\bibfield
  {journal} {\bibinfo  {journal} {{IOP} Conference Series: Materials Science
  and Engineering}\ }\textbf {\bibinfo {volume} {434}},\ \bibinfo {pages}
  {012034} (\bibinfo {year} {2018})}\BibitemShut {NoStop}%
\bibitem [{\citenamefont {Amit}(1989)}]{mbf}%
  \BibitemOpen
  \bibfield  {author} {\bibinfo {author} {\bibfnamefont {D.~J.}\ \bibnamefont
  {Amit}},\ }\href@noop {} {\emph {\bibinfo {title} {Modeling brain
  function}}}\ (\bibinfo  {publisher} {Cambridge University Press},\ \bibinfo
  {year} {1989})\BibitemShut {NoStop}%
\bibitem [{\citenamefont {{McEliece}}\ \emph {et~al.}(1987)\citenamefont
  {{McEliece}}, \citenamefont {{Posner}}, \citenamefont {{Rodemich}},\ and\
  \citenamefont {{Venkatesh}}}]{1057328}%
  \BibitemOpen
  \bibfield  {author} {\bibinfo {author} {\bibfnamefont {R.}~\bibnamefont
  {{McEliece}}}, \bibinfo {author} {\bibfnamefont {E.}~\bibnamefont
  {{Posner}}}, \bibinfo {author} {\bibfnamefont {E.}~\bibnamefont
  {{Rodemich}}}, \ and\ \bibinfo {author} {\bibfnamefont {S.}~\bibnamefont
  {{Venkatesh}}},\ }\href {\doibase 10.1109/TIT.1987.1057328} {\bibfield
  {journal} {\bibinfo  {journal} {IEEE Transactions on Information Theory}\
  }\textbf {\bibinfo {volume} {33}},\ \bibinfo {pages} {461} (\bibinfo {year}
  {1987})}\BibitemShut {NoStop}%
\bibitem [{\citenamefont {Gopalsamy}\ and\ \citenamefont
  {Liu}(2007)}]{GOPALSAMY2007375}%
  \BibitemOpen
  \bibfield  {author} {\bibinfo {author} {\bibfnamefont {K.}~\bibnamefont
  {Gopalsamy}}\ and\ \bibinfo {author} {\bibfnamefont {P.}~\bibnamefont
  {Liu}},\ }\href {\doibase https://doi.org/10.1016/j.nonrwa.2005.11.010}
  {\bibfield  {journal} {\bibinfo  {journal} {Nonlinear Analysis: Real World
  Applications}\ }\textbf {\bibinfo {volume} {8}},\ \bibinfo {pages} {375 }
  (\bibinfo {year} {2007})}\BibitemShut {NoStop}%
\bibitem [{\citenamefont {Singh}(2001)}]{PhysRevE.64.051912}%
  \BibitemOpen
  \bibfield  {author} {\bibinfo {author} {\bibfnamefont {M.~P.}\ \bibnamefont
  {Singh}},\ }\href {\doibase 10.1103/PhysRevE.64.051912} {\bibfield  {journal}
  {\bibinfo  {journal} {Phys. Rev. E}\ }\textbf {\bibinfo {volume} {64}},\
  \bibinfo {pages} {051912} (\bibinfo {year} {2001})}\BibitemShut {NoStop}%
\bibitem [{\citenamefont {Tsuboshita}\ and\ \citenamefont
  {Okada}(2010)}]{tsuboshita2010statistical}%
  \BibitemOpen
  \bibfield  {author} {\bibinfo {author} {\bibfnamefont {Y.}~\bibnamefont
  {Tsuboshita}}\ and\ \bibinfo {author} {\bibfnamefont {M.}~\bibnamefont
  {Okada}},\ }\href@noop {} {\bibfield  {journal} {\bibinfo  {journal} {journal
  of the Physical Society of Japan}\ }\textbf {\bibinfo {volume} {79}},\
  \bibinfo {pages} {024002} (\bibinfo {year} {2010})}\BibitemShut {NoStop}%
\bibitem [{\citenamefont {Bharitkar}\ and\ \citenamefont
  {Mendel}(2000)}]{bharitkar2000hysteretic}%
  \BibitemOpen
  \bibfield  {author} {\bibinfo {author} {\bibfnamefont {S.}~\bibnamefont
  {Bharitkar}}\ and\ \bibinfo {author} {\bibfnamefont {J.~M.}\ \bibnamefont
  {Mendel}},\ }\href@noop {} {\bibfield  {journal} {\bibinfo  {journal} {IEEE
  Transactions on neural networks}\ }\textbf {\bibinfo {volume} {11}},\
  \bibinfo {pages} {879} (\bibinfo {year} {2000})}\BibitemShut {NoStop}%
\bibitem [{\citenamefont {McClelland}\ and\ \citenamefont
  {Rumelhart}(1988)}]{McClelland1988}%
  \BibitemOpen
  \bibfield  {author} {\bibinfo {author} {\bibfnamefont {J.~L.}\ \bibnamefont
  {McClelland}}\ and\ \bibinfo {author} {\bibfnamefont {D.~E.}\ \bibnamefont
  {Rumelhart}},\ }\href {\doibase 10.3758/BF03203842} {\bibfield  {journal}
  {\bibinfo  {journal} {Behavior Research Methods, Instruments, {\&}
  Computers}\ }\textbf {\bibinfo {volume} {20}},\ \bibinfo {pages} {263}
  (\bibinfo {year} {1988})}\BibitemShut {NoStop}%
\bibitem [{\citenamefont {Hertz}\ \emph {et~al.}(1991)\citenamefont {Hertz},
  \citenamefont {Krogh},\ and\ \citenamefont {Palmer}}]{greenBook}%
  \BibitemOpen
  \bibfield  {author} {\bibinfo {author} {\bibfnamefont {J.}~\bibnamefont
  {Hertz}}, \bibinfo {author} {\bibfnamefont {A.}~\bibnamefont {Krogh}}, \ and\
  \bibinfo {author} {\bibfnamefont {R.~G.}\ \bibnamefont {Palmer}},\
  }\href@noop {} {\emph {\bibinfo {title} {Introduction to the theory of neural
  computation}}}\ (\bibinfo  {publisher} {Addison-Wesley Pub. Co.},\ \bibinfo
  {year} {1991})\BibitemShut {NoStop}%
\end{thebibliography}%

\end{document}